\documentclass[journal]{IEEEtai}

\usepackage[colorlinks,urlcolor=blue,linkcolor=blue,citecolor=blue]{hyperref}

\usepackage{color,array}

\usepackage{graphicx}
\usepackage{amsmath,amssymb,amsfonts}%
\usepackage{amsthm}%
\usepackage{mathrsfs}%

\begin{document}

\title{Recent advances in universal text embeddings: A Comprehensive Review of Top-Performing Methods on the MTEB Benchmark}

\author{Hongliu CAO, Amadeus SAS France}


\maketitle

\begin{abstract}
Text embedding methods have become increasingly popular in both industrial and academic fields due to their critical role in a variety of natural language processing tasks. The significance of universal text embeddings has been further highlighted with the rise of Large Language Models (LLMs) applications such as Retrieval-Augmented Systems (RAGs). While previous models have attempted to be general-purpose, they often struggle to generalize across tasks and domains. However, recent advancements in training data quantity, quality and diversity; synthetic data generation from LLMs as well as using LLMs as backbones encourage great improvements in pursuing universal text embeddings. In this paper, we provide an overview of the recent advances in universal text embedding models with a focus on the top performing text embeddings on Massive Text Embedding Benchmark (MTEB). Through detailed comparison and analysis, we offer a systematic organization of the literature, underscoring the significant developments and limitations in the recent advancements of universal text embedding models, and suggest potential future research directions that could inspire further advancements in this field.
\end{abstract}

\section{Introduction}\label{sec1}

Text embedding methods have gained considerable interest in both industry and academia due to their important role in various natural language processing tasks such as text classification \cite{li2021merging}, text clustering \cite{xu2023contrastive,sbertreimers2019sentence},  sentiment analysis \cite{suresh2021not,zhang2022leveraging}, information retrieval \cite{rajapakse2023dense}, question answering \cite{yue2021contrastive}, dialogue systems \cite{long2022multi}, semantic textual similarity \cite{grill2020bootstrap}, item recommendation \cite{steck2024cosine} and so on \cite{choi2021evaluation,gteli2023towards,wang2023improving}.
With the increasing popularity of Large Language Models (LLMs) based applications such as Retrieval-Augmented Systems (RAGs), the pivotal role of text embeddings has been underscored recently. This is mainly due to the fact that these LLM based applications are heavily dependent on the high quality text embeddings for tasks like vector search, a process where the most relevant documents are retrieved for LLM Question Answering (QA) \cite{gteli2023towards,asai2023retrieval}. 
Source attribution of generated text is another important application of text embeddings \cite{gao2023enabling} that can improve the interpretability and trustworthiness of LLMs \cite{e5mistralwang2023improving}.

\begin{figure*}
\includegraphics[width=\textwidth]{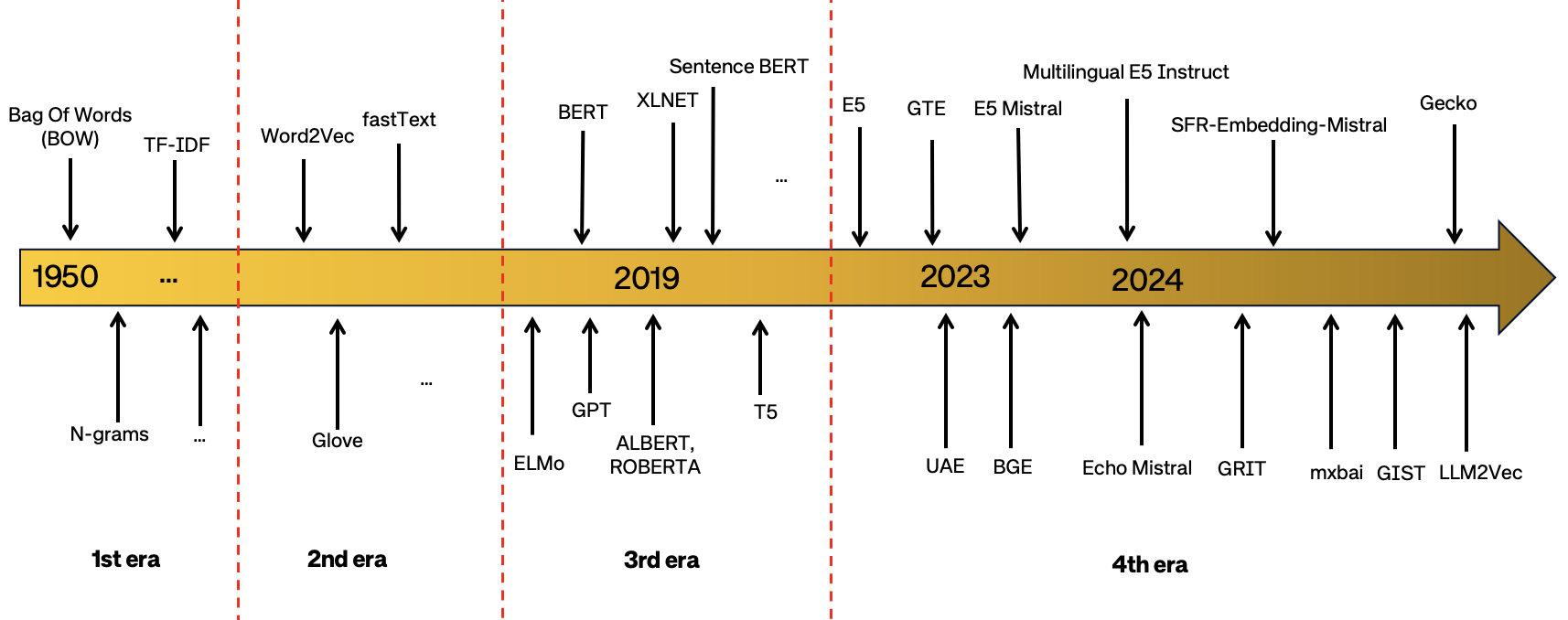}
\caption{The 4 different eras of text embeddings. 1st era: Count-based Embeddings (with dimension reduction techniques); 2nd era: Static dense word embeddings, 3rd era: Contextualized embeddings; 4th era: Universal text embeddings.} \label{timeline}
\end{figure*}

The field of text embeddings in natural language processing (NLP) has experienced significant changes over the past few decades. The shift from basic task specific representations to complex universal embeddings highlights the progress in this area (as shown in Figure \ref{timeline}):
\begin{itemize}
    \item 1st era: Count-based Embeddings.  Bag of Words  and Term Frequency-Inverse Document Frequency (TF-IDF) are two representative works in this era. Bag of Words \cite{harris1954distributional} is one of the earliest text representation methods, which counts the presence or occurrence of each word in the documents and use them as features.
    TF-IDF measures how important a word is to a document relative to a corpus, by increasing proportionally to the number of times a word appears in the document but offset by the frequency of the word in the corpus \cite{manning2008introduction}. Both BoW and TF-IDF highlight the word/term relevancy instead of using the context information or the meaning of words \cite{petukhova2024text}. There are also other works in this era transforming texts into low-dimensional dense embeddings such as  Latent Semantic Indexing (LSA) \cite{deerwester1990indexing} generating document embeddings with the decomposition of a word-document co-occurrence matrix \cite{e5wang2022text}. 
    
    \item 2nd era: Static dense word embeddings. Word2Vec \cite{word2vec}, GloVe \cite{glove} and FastText \cite{fasttext} are representative works that showed a significant step forward in the field of text representations using the surroundings of words to generate dense vector representations. Word2Vec focuses on local context using either Continuous Bag of Words (CBOW) approach (given the context, it predicts the target word) or Skip-gram approach (given the word, it predicts the context). Instead of only focusing on local context like Word2Vec, GloVe also takes the global corpus statistics into account. FastText further improves word embeddings by capturing the internal structure or morphology of words with a focus on character-level information of words and learning representations of sub-words \cite{patil2023survey}. Even though these models can capture a range of semantic and syntactic similarities successfully, they provide a single static vector per word, which ignores the fact that a word's meaning can be influenced by its surrounding context.
    
    \item 3rd era: Contextualized embeddings. The third era of text embeddings ushers in a new phase of embedding sophistication: context-sensitive dynamic embeddings that adapt or change based on context. Representative works include Embeddings From Language Models (ELMo) \cite{elmoneumann2018deep}, Generative Pre-trained Transformer (GPT) \cite{gpt1radford2018improving} and Bidirectional Encoder Representations from Transformers (BERT) \cite{devlin2018bert}. ELMo models the polysemy using a bidirectional Long Short Term Memory Network (LSTM) with the concatenation of the left-to-right and right-to-left representations.  Unlike ELMo, GPT uses Transformer (one-way instead of bi-directional) \cite{vaswani2017attention} to learn the text embedding using a combination of unsupervised pre-training and supervised fine-tuning. It was observed that attentional memory of the transformer assisted in (better) transfer compared to LSTMs \cite{patil2023survey}. BERT instead uses a bidirectional Transformer encoder to take into account both the left and right context for Masked Language Model (MLM) and Next Sentence Prediction (NSP) tasks during pre-training, which allows for a deeper understanding of word relationships by considering the full context of a word in a sentence in both directions. \cite{devlin2018bert,clusterpetukhova2024text}.  
    
    \item 4th era: Universal text embeddings.  The pursuit of developing a unified model to address a multitude of downstream tasks has been long-standing \cite{gteli2023towards}. Despite attempting to be general-purpose in previous models such as \cite{cer2018universal,t5raffel2020exploring,st5ni2021sentence}, studies indicate that these embedding models struggle to generalize across tasks and domains \cite{lee2024gecko}. Thanks to the increasing number and improved quality of diverse text datasets across different tasks \cite{bgexiao2023c,asai2022task}, good quality synthetic data generated by LLMs \cite{lee2024gecko,e5mistralwang2023improving} as well as benchmarks with the focus on novel task and domain generalization such as the Massive Text Embedding Benchmark (MTEB) \cite{muennighoff2022mteb}; the universality of text embeddings can be improved and evaluated across various languages and tasks such as retrieval, ranking, clustering, among others.
    The creation of unified models trained across diverse tasks has started to make progress with representative works like GTE \cite{gteli2023towards}, BGE \cite{bgexiao2023c}, E5 \cite{e5wang2022text,e5mistralwang2023improving,e5instructwang2024multilingual}, Gecko \cite{lee2024gecko}, LLM2Vec \cite{llm2vec}, etc.

\end{itemize}

There are several reviews on text embeddings such as in \cite{camacho2018word,ruder2019survey,patil2023survey,selva2021review,liu2020survey,kashyap2024reviewcomprehensive}, but none of the existing work focus on the recent advances in the universal text embeddings in the fourth era. To fill in the gap, the main focus of this work is to review recent advances in universal text embeddings. More specifically, the top performing text embeddings in the Massive Text Embedding Benchmark (MTEB)  \cite{muennighoff2022mteb} are the main focus of this work.   
This survey offers a systematic organization of the literature, underscoring the significant developments and challenges in the recent advancements of universal text embedding models (primarily focusing on the methods proposed in years 2023 and 2024). Furthermore, we suggest potential future research directions that could inspire further advancements in this field.
The reminder of this paper is organized as follows: the preliminaries, background and categorization of 4th era universal text embeddings are introduced in Section 2. In Section 3, 4 and 5, the overview of the top performing state of the art text embeddings and their main contributions are explained. We describe the trends, performance and efficiency analysis of the state of the art text embeddings as well as their limitations in Section 6.   Finally, the conclusion and future directions in text embeddings are given in Section 7.

\section{Preliminaries}

\subsection{Definitions}
\paragraph{\textbf{Text embedding}}
In the context of Natural Language Processing (NLP) or  Natural Language Understanding (NLU), text refers to a collection of words, phrases, sentences, paragraphs or larger utterance that convey meaningful information \cite{indurkhya2010handbook}. The form and length of text often vary on the task such as text classification/clustering,  sentiment analysis, information retrieval, dialogue systems, item recommendation, etc. However, an embedding is a fixed-length low-dimensional dense vector representation \cite{e5wang2022text}. Text embedding then can be defined as a numerical dense representation of a word, phrase, sentence, or larger utterance in natural language in a certain space where texts with similar meanings are near each other \cite{sbertreimers2019sentence,lee2024gecko,liu2020survey,gteli2023towards}. The meaning of a word is  influenced by its context, and it is from this context that a word embedding is usually learnt. The meaning of a sentence is more complex because it depends on the words used in the sentence, the  syntactic structure as well as the surrounding sentences \cite{li2022brief}.  The meaning of a document is even more complex as it is a high-level abstraction of the whole text (words, sentences, paragraphs, etc.). The definition of "meaning", "local information" or "context" changes when the text length changes, which makes it a great challenge to learn the embedding for an "arbitrary span of contiguous text" \cite{devlin2018bert}.

\paragraph{\textbf{Universal text embedding}} In recent works, universal text embedding \cite{bgexiao2023c,llm2vec,gistsolatorio2024gistembed} or general-purpose text embedding as used in \cite{lee2024gecko,gteli2023towards,e5wang2022text,echoispringer2024repetition} generally means  a unified comprehensive text embedding model that can address a multitude of downstream tasks. In other words, the universal text embedding is not just proficient in a single particular task, but it proves to be consistently beneficial across a range of tasks such as text classification, text clustering, sentiment analysis, semantic textual similarity, summarization, retrieval tasks, etc. The objective of creating universal text embeddings is to mimic the fundamental process of how humans understand and process text, which can be beneficial in various domains \cite{li2022brief}. With the recent work such as \cite{e5instructwang2024multilingual,gritmuennighoff2024generative}, the definition of universal text embedding has been extended to multi-task, multi-lingual, while \cite{gteli2023towards} shows that a natural language model can also understand well programming languages. In this work, we define universal text embedding as \textbf{  a unified comprehensive text embedding model that can address a multitude of input text length, downstream tasks, domains and languages}. 
The research of universal text embedding has been stimulated by several recent developments. These include the growth in quantity and refinement in quality of diverse text datasets across various tasks \cite{bgexiao2023c,asai2022task}, the production of high-quality synthetic data by LLMs \cite{lee2024gecko,e5mistralwang2023improving}, and benchmarks that emphasize new task and domain generalization, such as the multi-lingual Massive Text Embedding Benchmark (MTEB) \cite{muennighoff2022mteb}.


\subsection{Background}
In this work, we study and analyze the top performing text embedding models that are either open-source or well documented from MTEB English benchmark (because the English benchmark has more and diverse evaluation tasks compared to other languages). It can be found that 
BERT-based models used in \cite{gteli2023towards,bgexiao2023c,e5wang2022text,e5instructwang2024multilingual,li2023angle} and LLMs used in \cite{e5instructwang2024multilingual,gritmuennighoff2024generative,llm2vec,SFRAIResearch2024,echoispringer2024repetition,lee2024gecko} are two most popular backbones of the top performing universal text embedding models on the MTEB English benchmark. 

\paragraph{\textbf{BERT} \cite{devlin2018bert}} To generate contextual embeddings, BERT, pre-trained on a massive corpus and fine-tuned using labeled data from the downstream tasks, employs a bidirectional Transformer encoder to take into account both the left and right context in all layers.  To alleviate the uni-directionality constraint, BERT proposes a masked language modelling (MLM) objective, where some of the tokens of a input sequence are randomly masked, and the objective is to predict the vocab-ids of the masked tokens based only on its context \cite{devlin2018bert}. Additionally, a Next Sentence Prediction (NSP) task is also used to jointly pre-train text-pair representations with the objective to help tasks that require reasoning over text pairs \cite{liu2020survey}. 
WordPiece embeddings with a 30,000 tokens vocabulary \cite{wu2016google} is used by BERT, with special tokens including [CLS] token (a special classification token as the first token of each sequence) and [SEP] token to separate sentence pairs.   
The final hidden state of [CLS] is used for sentence-level tasks and the final hidden state of each token is used for token-level tasks \cite{devlin2018bert,liu2020survey}. Some important details about BERT include:
\begin{itemize}
    \item Pre-training data: BooksCorpus (800M words) \cite{zhu2015aligning} and English Wikipedia ignoring lists, tables, and headers (2,500M words). 
    \item Fine-tuning: task-specific inputs and outputs are fed into BERT to Fine-tuning all the parameters end-to-end. 
    \item Loss function: the sum of the mean MLM likelihood and the mean NSP likelihood \cite{devlin2018bert}.
    \item Model size: $BERT_{BASE}: 110M, BERT_{LARGE}: 340M$. 
    \item Training: Training of $BERT_{BASE}$ was performed on 4 Cloud TPUs in Pod configuration (16 TPU chips total). Training of $BERT_{LARGE}$ was performed on 16 Cloud TPUs (64 TPU chips total). Each pre-training took 4 days to complete.
\end{itemize}

Following the success of BERT, several BERT-based models have been introduced, such as Robustly Optimized BERT Pretraining Approach (RoBERTa) \cite{liu2019roberta}, Distilled version of BERT (DistilBERT) \cite{sanh2019distilbert}, and A Lite BERT (ALBERT) \cite{lan2019albert}, each offering unique enhancements and optimizations while maintaining the core bidirectional approach of the original BERT model. One of the limitations of the BERT network structure is that no independent sentence embeddings are computed, which makes it difficult to use for various pair regression tasks due to large number of combinations.  To allow for more efficient sentence-level embeddings, Sentence-BERT (SBERT) introduces the siamese and triplet network structures to  generate highly effective semantically meaningful sentence embeddings that can be compared with cosine similarity, which has served as a cornerstone for further research \cite{sbertreimers2019sentence,kashyap2024reviewcomprehensive}. Another cornerstone work is Simple Contrastive Learning of Sentence Embeddings (SimCSE) \cite{nligao2021simcse} using unsupervised and supervised contrastive learning, which is widely adopted by recent state of the art text embeddings.

\paragraph{\textbf{Large Language Models}}
The widespread use of ChatGPT has showcased the impressive abilities of Large Language Models (LLMs) in following instructions, in-context learning with minimal few-shot examples and amazing conversation
abilities with humans. While some of the best performing LLMs like GPT-4 \cite{achiam2023gpt} are proprietary with limited technical information available,  some open-source LLM  models like LLaMA-2 \cite{touvron2023llama}, LLaMA-3 \cite{llama3modelcard}  and Mistral \cite{jiang2023mistral} have made some notable efforts to catch up\cite{e5mistralwang2023improving}.  One advantage of using LLMs for text embedding is that they are extensively pre-trained on web-scale data already, which does not need the contrastive pre-training step used in existing state of the art text embedding models.  At present, the foundation for the majority of LLMs is the Transformer architecture, which employs layers of multi-head attention in a very deep neural network. Decoder-only LLMs utilize the causal attention mechanism, where the representation of a token at a specific position $i$ is exclusively impacted by the representations of tokens that come before it. The authors from \cite{llm2vec} hypothesize that  causal attention mechanism might partly be the  reason of the slow adoption of decoder-only LLMs for text embedding tasks as it inherently limits their ability to produce rich contextualized representations. Several recent works such  \cite{e5instructwang2024multilingual,gritmuennighoff2024generative,llm2vec,SFRAIResearch2024,echoispringer2024repetition,lee2024gecko} have proposed several solutions to mitigate such limitations.

\paragraph{\textbf{Massive Text Embedding Benchmark (MTEB)}}
The objective of MTEB is to provide comprehensive understandings on the universality of  text embedding models, including 58 datasets covering 112 languages from 8 embedding tasks: Bitext mining, Classification, Clustering, Pair classification, Reranking, Retrieval, Semantic Textual Similarity (STS) and Summarization \cite{muennighoff2022mteb}. The leader-board results are available on the Hugging Face Hub\footnote{https://huggingface.co/spaces/mteb/leaderboard}, where the results of English (56 datasets), Chinese (35 datasets), French (26 datasets) and Polish (26 datasets) benchmark results can be found respectively.

\subsection{Taxonomy of universal text embeddings}
\begin{figure}
\includegraphics[width=0.5\textwidth]{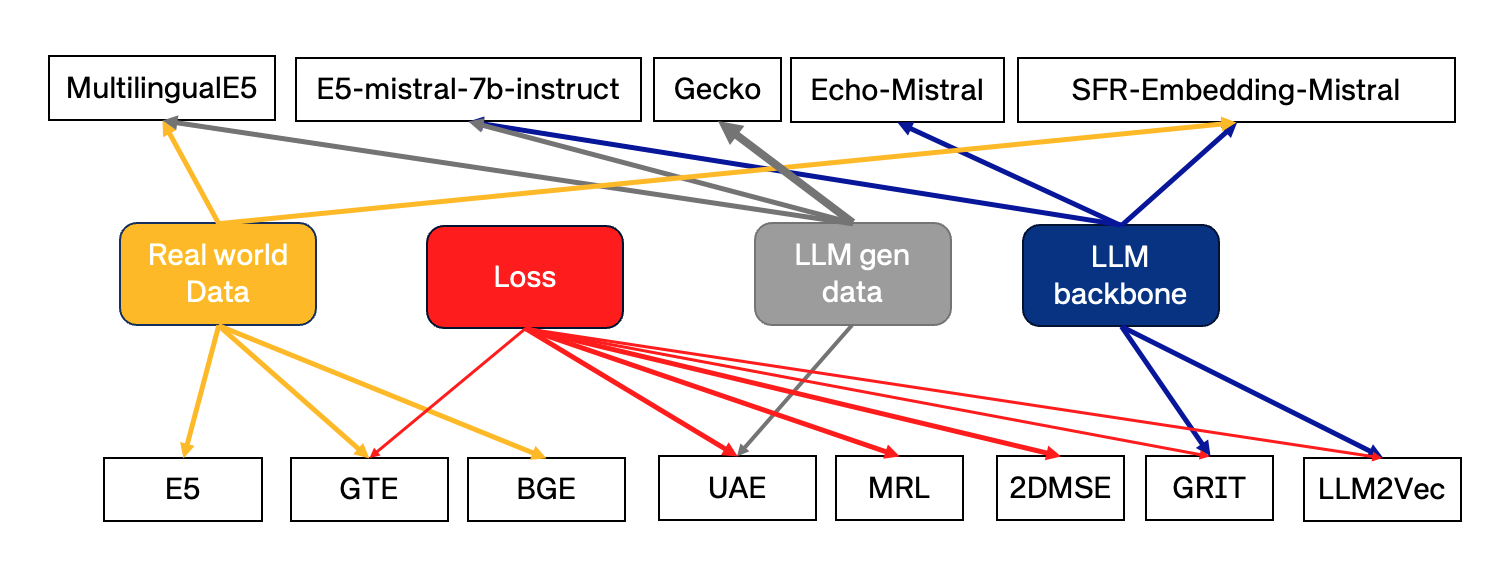}
\caption{Representative state of the art universal text embeddings and their main focus/contributions.} \label{taxo}
\end{figure}

In this section, the main focuses and contributions of the some of the MTEB top performing state of the art text embedding methods are analyzed (shown in Figure \ref{taxo}), including:
 E5: EmbEddings from bidirEctional Encoder rEpresentations  \cite{e5wang2022text},
GTE: General-purpose Text Embedding model \cite{gteli2023towards},
BGE: Beijing Academy of Artificial Intelligence (BAAI) General Embedding \cite{bgexiao2023c},
UAE: Universal AnglE Embedding \cite{li2023angle},
MRL: Matryoshka Representation Learning \cite{mrlkusupati2022matryoshka}, 
2DMSE: 2D Matryoshka Sentence Embeddings  \cite{2dmrlli20242d},
 GRIT: Generative Representational Instruction Tuning \cite{gritmuennighoff2024generative},
 LLM2Vec: \cite{llm2vec}, 
Multilingual E5: \cite{e5instructwang2024multilingual},
E5-mistral-7b-instruct: \cite{e5mistralwang2023improving}, 
 Gecko: \cite{lee2024gecko},
 Echo-mistral: \cite{echoispringer2024repetition},
SFR-Embedding-Mistral: \cite{SFRAIResearch2024}.
The main focus/contributions are summarized and simplified as the following 4 aspects: 
\begin{itemize}
    \item Real world data: one way to learn the universal text embedding is using  a multi-stage contrastive learning strategy with diverse training data mixture. For example, GTE \cite{gteli2023towards} uses diverse datasets for both pre-training and fine-tuning stage. BGE \cite{bgexiao2023c} introduces a compressive data package C-Pack, while E5 \cite{e5wang2022text} constructed a curated web-scale text pair dataset  named Colossal Clean text Pairs (CCPairs) containing heterogeneous training signals by combining various semi-structured data sources along with aggressive filtering (270M text pairs filtered from 1.3B noisy text pairs) with a consistency-based filter to improve data quality \cite{dai2022promptagator}. Some works like GISTEmbed \cite{gistsolatorio2024gistembed} also focus on improving the quality of hard negatives used for training.
    \item Loss function: another research direction is to focus on improving the loss functions. As many existing text embedding works employed the cosine function in their training objective to measure the pairwise semantic similarity, the authors from UAE \cite{li2023angle} point out that there is the gradient vanishing issue due to the saturation zones of cosine function, which hinder the ability to learn subtle distinctions between texts in back propagation. Hence they propose a novel angle-optimized text embedding model called AnglE with angle optimization in a complex space which substantially improve the text embedding quality in various scenarios.  Matryoshka Representation Learning (MRL) \cite{mrlkusupati2022matryoshka} and 2D Matryoshka Sentence Embeddings (2DMSE) propose new loss functions in order to reduce the computational cost of downstream tasks. 
    \item LLMs are used to improve the universal text embeddings in two different ways: 
    \begin{itemize}
        \item 1. use synthetic data generated by LLMs: In \cite{li2023angle}, the authors apply LLMs as data annotators to label the pseudo-supervised data for the training to improve the model performance. \cite{e5mistralwang2023improving} and \cite{e5instructwang2024multilingual} use proprietary LLMs including GPT-35-Turbo and GPT-4 to generate  synthetic data covering a various range of text embedding tasks in 93 languages (among which 25\% are generated by GPT-35-Turbo and others are generated by GPT-4) to increase the training data diversity, while \cite{lee2024gecko} use synthetic data generation to distill knowledge from large language models into a retriever.
        \item 2. use LLMs as backbone for text embeddings: as LLMs are extensively pre-trained on web-scale data already, which does not need the large scale contrastive pre-training step used in existing state of the art text embedding models, many works also try to get embeddings directly from LLMs. For example, E5-mistral-7b-instruct perform multi-task fine-tuning on Mistral 7b model which is one of the best performing method on MTEB. Echo-mistral \cite{echoispringer2024repetition}, LLM2Vec \cite{llm2vec} and GRIT \cite{gritmuennighoff2024generative} propose various different ways so that decoder only LLMs can generate high quality text embeddings using bidirectional attention. 
    \end{itemize}
     
\end{itemize}

From Figure \ref{taxo}, it can be seen that most works have multiple contributions. To make the taxonomy easier, the state of the art text embeddings are divided into 3 groups based on their main contributions/focuses: Data focused text embeddings (detailed in Section 3), Loss focused text embeddings (detailed in Section 4) and LLM focused text embeddings (detailed in Section 5).

\section{Data focused universal text embeddings}
One way to learn the universal text embedding is using a multi-stage contrastive learning strategy with improved training data mixture in terms of data quantity, quality and diversity as summarized in Table \ref{data}. For example, GTE \cite{gteli2023towards} uses diverse datasets for both pre-training and fine-tuning stage. BGE \cite{bgexiao2023c} introduces a compressive data package C-Pack, while E5 \cite{e5wang2022text} constructed a curated web-scale text pair dataset  named Colossal Clean text Pairs (CCPairs) containing heterogeneous training signals by combining various semi-structured data sources along with aggressive filtering (270M text pairs filtered from 1.3B noisy text pairs) with a consistency-based filter to improve data quality \cite{dai2022promptagator}. Some works like GISTEmbed \cite{gistsolatorio2024gistembed} also focus on improving the quality of hard negatives used for training. More details about each text embedding methods can be found below. 

\begin{table}[t!]
\caption{The main contributions of data focus universal text embeddings: quantity, quality and diversity.}\label{data}
 \centering
\begin{tabular}{ p{1.5cm}  p{6cm}}
\\  \hline
Model names & Data focus contribution \\
 \hline
GTE \cite{gteli2023towards} & Substantial performance gains are achieved by notably augmenting \textbf{ data volume} during both unsupervised pre-training (800M text pairs used for pre-training) and supervised fine-tuning stages with \textbf{ diverse mixture} of datasets from multiple sources.  \\ \hline

BGE \cite{bgexiao2023c} & The \textbf{largest} dataset C-MTP was developed for general Chinese embedding with the focuses on:  1. \textbf{data quality} improvement by filtering the irrelevant text pairs in unlabelled data for general purpose fine-tuning (around 100M text pairs); 2. \textbf{multi-task} high quality labelled data (838,465 text pairs) for task-specific fine-tuning.
Note: English data (for English version of BGE) is 2 times larger than the Chinese data.   \\ \hline

GISTEmbed \cite{gistsolatorio2024gistembed} & GIST is fine-tuned on top of BGE on MEDI and MTEB classification datasets with improved \textbf{in-batch negative data quality}.  \\  \hline

E5 \cite{e5wang2022text} & Development of CCPairs: curated \textbf{large-scale} text pair dataset by harvesting \textbf{heterogeneous} semi-structured data sources using consistency-based filter for \textbf{quality improvement} (reducing 1.3B text pairs to 270M text pairs for pre-training). \\  \hline

Multilingual-E5 \cite{e5instructwang2024multilingual} & \textbf{Multilingual} focus: diverse mixture of multilingual text pairs obtained from various sources (1B text pairs). Additional 500k \textbf{synthetic data} generated by GPT-3.5/4 which encompasses 150k unique instructions and covers 93 languages were used for fine-tuning. \\

 \hline
\end{tabular}
\end{table}

\paragraph{\textbf{ General-purpose Text Embedding model (GTE) }} With the focus on developing a unified more comprehensive model for general text representation  to address a multitude of downstream tasks, the authors from \cite{gteli2023towards} introduce a multi-stage contrastive learning strategy with diverse training data mixture:  in the initial stage,  a large corpus of open-source data without any filtering or cleaning are used to learn basic language patterns with unsupervised contrastive learning; in the second stage, supervised fine-tuning refines the embeddings using contrastive learning with a smaller, high-quality dataset. At both stages,  the number of training data are significantly increased.

For a query $q$, a relevant/positive document $d^+$, a set of irrelevant/negative documents $D_{-} = \{ d_{-}^{1} , . . . , d_{-}^{n} \} $, the InfoNCE loss \cite{infonceoord2018representation} is defined as in Equation \ref{infonce}:
\begin{equation}
\label{infonce}
    \mathcal{L}_{cl} = -log \frac{e^{s(q, d^+)/\tau }}{e^{s(q, d^+)/\tau} + \sum_{i=1}^n e^{s(q, d^-)/\tau}}
\end{equation}
where $s(q, d)$ estimates the similarity between two pieces of text $q$ and $d$ via vector distance between their embeddings $q = E(q)$ and $d = E(d)$.

In GTE, given a batch of positive text pair samples $\{ (q_1, d_1), (q_2, d_2), ..., (q_n, d_n)  \}$, the authors propose an improved contrastive loss (icl) can be viewed as a combination of loss variants proposed by \cite{radford2021learning,ren2021pair,moiseev2023samtone}:
\begin{equation}
\label{icl}
    \mathcal{L}_{icl} = -\frac{1}{n} \sum_{i=1}^n log  \frac {e^{s(q_i, d_i)/\tau} }{Z}  
\end{equation}
where 
\begin{equation}
\label{iclz}
    Z =  \sum_{j} e^{s(q_i, d_j)/\tau} + \sum_{j \neq i} e^{s(q_i, q_j)/\tau} +  \sum_{j} e^{s(q_j, d_i)/\tau} + \sum_{j \neq i} e^{s(d_j, d_i)/\tau} 
\end{equation}

The cosine similarity is used as the similarity measure $s(q, d)$.
GTE models are initialized with pre-trained language models such as BERT with mean pooling on top of the contextualized token representations produced by the language model for text embeddings.
Some other details about GTE include:
\begin{itemize}
    \item Pre-training data: around 800M text pairs text pairs for the unsupervised pre-training (a multinomial distribution is used to sample data batches from different data sources, taking into account their respective sizes.):
    \begin{itemize}
            \item{Web page (147M)}: Common Crawl, Clue Webs, MS MARCO documents, title as query and the body text as document.
            \item{Academic Paper (45M)}: arXiv, bioRxiv, medRxiv, PubMed and Semantic Scholar, title as query and its abstract as document
            \item{Hyperlink (106M)}: ClueWeb, Wikipedia and Semantic Scholar paper citations, the citation argument and the text from reference as relevant text pairs for contrast.
            \item{Social Media (327M)}: Reddit, title  body pair, post comment pair
            \item{Knowledge Base (38M)}: WikiPedia and DBPedia, entity, description pairs
            \item{Community QA (12M)}: StackExchange, Yahoo Answers, WikiHow and Amazon QA, summaritive title and a descriptive body pairs and question answer pairs
            \item{News (3M)}: CCNews, MicrosoftNews, NPR, CNNDaily, title body pairs
            \item{Code (20M)}: GitHub (CodeSearchNet) and StackOverflow, text-code pairs
            \item{Others (91M)}: Amazon reviews about the goods, debate websites about one argument, googaq query answer pairs by prompting google search box with search log queries.
            
       \end{itemize}
    \item Fine-tuning data: 
    \begin{itemize}
            \item{Web Search}: MS MARCO \cite{bajaj2016ms} passage retrieval benchmarks where hard negatives are mined by sampling from high-ranked documents retrieval system, excluding positive ones.
            \item{Open QA}: Natural Questions (NQ), Trivia QA \cite{nq1karpukhin2020dense,nq2kwiatkowski2019natural}, Web Questions, HotpotQA \cite{yang2018hotpotqa}, etc. Top ranked passage by retrieval system which do not include answer to the question is regarded as hard negatives.
           \item{Natural Language Inference}: MNLI \cite{mnliwilliams2017broad} and SNLI \cite{snlibowman2015large}, entailment as positive pairs and contradiction as negative pairs 
           
           \item{Fact Verification}: training set from FEVER \cite{thorne2018fever}

           \item{Paraphrase}:  Quora \cite{iyer2017quora} and StackExchange Dupquestion
           \item{Others}: miscellaneous datasets from different NLP tasks and domains released in MEDI \cite{su2022one} and BERRI \cite{asai2022task}.
            
       \end{itemize}

    \item Loss function: improved contrastive loss as Equation \ref{icl}
    
    \item Negative sampling: 
    \begin{itemize}
        \item Pre-training: enlarged in-batch negatives,
        \item Fine-tuning: hard negatives mined by an extra retriever to form text triples. 
    \end{itemize}
     
    \item Model size: 
    \begin{itemize}
        \item $GTE_{small}$: 30M (backbone: MiniLM-L12-H384-uncased),
      \item  $GTE_{base}$: 110M (backbone: bert-base-uncased)  
        \item   $GTE_{large}$: 330M (backbone: bert-large-uncased)  
    \end{itemize}
      
\end{itemize}

\paragraph{\textbf{ Beijing Academy of Artificial Intelligence (BAAI) General Embedding (BGE) }} Similar to the objective of GTE, BGE also tries to learn general-purpose text embeddings, a comprehensive, unified embedding model which is capable of managing all types of uses, including retrieval, ranking, and classification, across various application settings such as question answering, language modeling, and conversation \cite{bgexiao2023c}.  BGE introduces C-Pack, a comprehensive package designed to advance the general Chinese embedding (other languages version of BGE are also available), along with their training recipe:  pre-training of an embedding-oriented text encoder, general-purpose contrastive learning, and task-specific fine-tuning.  BERT-like architecture is used by BGE models where the last layer’s hidden state of the special token [CLS] is trained to work as the embedding (unlike GTE). Another major difference from GTE is that BGE uses instruction-based fine-tuning to deal with potentially mutually contradicted tasks: a task specific instruction which describes the nature of the task (e.g. search relevant passages for the query) is added to the query side for each text pair. 
Some other details about BGE include:
\begin{itemize}
    \item Pre-training data (English version): unsupervised datasets including datasets like Wikipedia, CC-net, StackExchange, Reddit, S2ORC \cite{lo2019s2orc}, and datasets from sentence-transformers\footnote{https://huggingface.co/datasets/sentence-transformers/embedding-training-data}.
    \item Fine-tuning data (English version):  supervised datasets including NLI \cite{nligao2021simcse}, FEVER \cite{thorne2018fever}, NQ \cite{nq1karpukhin2020dense,nq2kwiatkowski2019natural}, HotpotQA \cite{yang2018hotpotqa}, Quora \cite{iyer2017quora}, StackExchange Duplicates and MEDI \cite{su2022one}.

    \item Loss function: the contrastive loss as in Equation \ref{infonce}
    \item Negative sampling: 
    \begin{itemize}
        \item Pre-training: purely rely on in-batch negative samples \cite{karpukhin2020dense} and resort to a big batch size (as large as 19,200) to improve the discriminativeness of the embedding.
        \item Fine-tuning:  in addition to the in-batch negative samples, one hard negative sample is mined for each text pair from the task’s original corpus, following the ANN-style sampling strategy in \cite{xiong2020approximate}
    \end{itemize}

    \item Model size: 
    \begin{itemize}
        \item $BGE_{small}$: 24M (BERT-like architecture),
      \item  $BGE_{large}$: 102M (BERT-like architecture), 
        \item   $BGE_{large}$: 326M (BERT-like architecture).  
    \end{itemize}
      
\end{itemize}
\textbf{ Guided In-sample Selection of Training Negatives for Text Embedding Fine-tuning (GISTEmbed)} GIST-large-Embedding-v0 is another top performing text embeddings on the MTEB benchmark which uses $BGE_{large}$ as backbone. The main focus of  GISTEmbed is to propose a novel strategy that enhances in-batch negative selection during contrastive training through a guide model \cite{gistsolatorio2024gistembed}, which improves the baseline performance slightly. However, the GIST-large-Embedding-v0 performance increase on MTEB benchmark compared to  $BGE_{large}$ is limited (0.11\%). It is difficult to analyze if the  limited performance increase is due to the proposed guided in-sample negative selection or due to the fact that they added in-domain MTEB training data to fine-tune the BGE embedding models. 

\paragraph{\textbf{ EmbEddings from bidirEctional Encoder rEpresentations (E5) }}  With the objective of creating high-quality general-purpose text embeddings suitable for any tasks requiring single-vector representations in both zero-shot or fine-tuned settings, the authors from \cite{e5wang2022text} constructed a curated web-scale text pair dataset  named Colossal Clean text Pairs (CCPairs) containing heterogeneous training signals by combining various semi-structured data sources such as CommunityQA, Common Crawl and Scientific papers along with aggressive filtering (270M text pairs filtered from 1.3B noisy text pairs) with a consistency-based filter to improve data quality \cite{dai2022promptagator}. 
Some other details about E5 include:
\begin{itemize}
    \item Pre-training data: (post, comment) pairs from Reddit\footnote{https://files.pushshift.io/reddit/}, (question, upvoted answer) pairs from Stackexchange\footnote{https://archive.org/details/stackexchange}, (entity name + section title, passage) pairs from English Wikipedia, (title, abstract) and citation pairs from Scientific papers \cite{lo2019s2orc}, and (title, passage) pairs from Common Crawl web pages\footnote{https://commoncrawl.org/}, various News sources and others including “SimpleWiki”, “GooAQ”, “WikiHow”, “Yahoo Answers” \footnote{https://huggingface.co/datasets/sentence-transformers/embedding-training-data}.
    \item Fine-tuning data: Natural Language Inference (NLI \cite{snlibowman2015large}), MS-MARCO passage ranking dataset \cite{bajaj2016ms}, and Natural Questions (NQ) dataset \cite{nq1karpukhin2020dense,nq2kwiatkowski2019natural}
    \item Loss function: 
    \begin{itemize}
        \item Pre-training: the contrastive loss as in Equation \ref{infonce}
        \item Fine-tuning:  a linear interpolation between contrastive loss for hard labels and KL divergence for distilling soft labels from the teacher model
    \end{itemize}  
    \item Negative sampling: 
    \begin{itemize}
        \item Pre-training: in-batch negative samples (with large 32,768 batch size)
        \item Fine-tuning: in-batch negative samples, mined hard negatives and knowledge distillation from a cross-encoder (CE) teacher model for the MS-MARCO and NQ datasets. For the NLI dataset, contradiction sentences are regarded as hard negatives.
    \end{itemize}
    \item Model size: 
    \begin{itemize}
        \item $E5_{small}$: 33M, initialized from MiniLM \cite{wang2020minilmv2}
        \item $E5_{base}$: 110M, initialized from bert-base-uncased
        \item $E5_{large}$: 330M, initialized from bert-large-uncased-whole-word-masking
    \end{itemize}
      
\end{itemize}

\paragraph{\textbf{ Multilingual-E5}} 

In order to extend the English E5 models, the authors from \cite{e5instructwang2024multilingual} released Multilingual-E5 series models by using diverse mixture of multilingual text pairs obtained from various sources with around 1 billion text pairs.
The English E5 model recipe is used for the training procedure, which involves contrastive pre-training on 1 billion multilingual text pairs and fine-tuning on a blend of labeled datasets, with the Multilingual-E5-large-instruct adopting the data mixture from \cite{e5mistralwang2023improving} that includes an additional 500k synthetic data created by GPT-3.5/4 and encompasses 150k unique instructions across 93 languages. Similar to BGE, instructions data are used to better inform embedding models about the task at hand for Multilingual-E5-large-instruct model.
Some other details about Multilingual-E5 include:
\begin{itemize}
    \item Pre-training data: around 1 billion multilingual text pairs from: Wikipedia, mC4, Multilingual CC News, NLLB, Reddit, S2ORC, Stackexchange, xP3 and Misc. SBERT Data.
    \item Fine-tuning data: blend of labeled datasets (around 1.6M) from MS-MARCO Passage, MS-MARCO Document NQ, TriviaQA, SQuAD, NLI, ELI5, NLLB, DuReader Retrieval, Fever, HotpotQA, Quora Duplicate Questions, Mr. TyDi and MIRACL (additional synthetic data with 150k unique instructions and covers 93 languages are used for fine-tuning  Multilingual-E5-large-instruct model).
    \item Loss function: 
    \begin{itemize}
        \item Pre-training: the contrastive loss as in Equation \ref{infonce}
        \item Fine-tuning:  a linear interpolation between contrastive loss for hard labels and KL divergence for distilling soft labels from the teacher model
    \end{itemize}  
    \item Negative sampling: 
    \begin{itemize}
        \item Pre-training: in-batch negative samples
        \item Fine-tuning: in-batch negative samples, mined hard negatives and knowledge distillation from a cross-encoder (CE) teacher model.
    \end{itemize}
    \item Model size: 
    \begin{itemize}
        \item Multilingual-E5-small: 118M (initialized from multi-lingual MiniLM \cite{wang2020minilmv2}),
      \item  Multilingual-E5-base: 278M (initialized from xlm-roberta-base \cite{conneau2019unsupervised}), 
      \item Multilingual-E5-large:  560M (initialized from xlm-roberta-large \cite{conneau2019unsupervised})
      \item Multilingual-E5-large-instruct: 560M, fine-tuned with instruction data on Multilingual-E5-large.
    \end{itemize}
      
\end{itemize}

\paragraph{\textbf{Summary}} In this section, the state of the art methods trying to achieve universal text embeddings with improved data quantity, quality and diversity are introduced. Most of these methods use  datasets from Common Crawl
, Wikipedia, social media, academic papers and  sentence-transformers\footnote{https://huggingface.co/datasets/sentence-transformers/embedding-training-data} (fully or partially) as one part of the pre-training data.  Code data and hyperlinks are also used by GTE, which enables GTE to understand both natural language and code. Similarly, Multilingual-E5 improve the data diversity by adding both real world and synthetic multilingual datasets in order to improve the universality across languages. On the other hand, most of these methods use hard negatives to improve the quality of negative samples. GISTEmbed proposes in-batch negative selection for better negative samples. E5 uses preliminary filters and consistency based filter to improve the training data quality while reducing pre-training data size from 1.3B to 270M. High quality multi-task datasets are also used by most of the methods during fine-tuning stage to improve the universality across downstream tasks. 

\section{Loss focused universal text embeddings}
Contrastive learning with InfoNCE loss (Equation \ref{infonce}) is used by most of the state of the art universal text embedding models. Several loss variants have been proposed by \cite{radford2021learning,ren2021pair,moiseev2023samtone} and  the authors of GTE propose an improved contrastive loss (Equation \ref{icl}) which combines these loss variants.  
As many existing text embedding works employed the cosine function in their training objective to measure the pairwise semantic similarity, the authors from UAE \cite{li2023angle} point out that there is the gradient vanishing issue due to the saturation zones of cosine function, which hinder the ability to learn subtle distinctions between texts in back propagation. Hence they propose a novel angle-optimized text embedding model called AnglE with angle optimization in a complex space which substantially improve the text embedding quality in various scenarios.  Matryoshka Representation Learning (MRL) \cite{mrlkusupati2022matryoshka} and 2D Matryoshka Sentence Embeddings (2DMSE) propose new loss functions in order to reduce the computational cost of downstream tasks. More details about different novel losses proposed by the MTEB top performing universal text embedding models can be found below.

\begin{figure}
\centering
\includegraphics[width=0.45\textwidth]{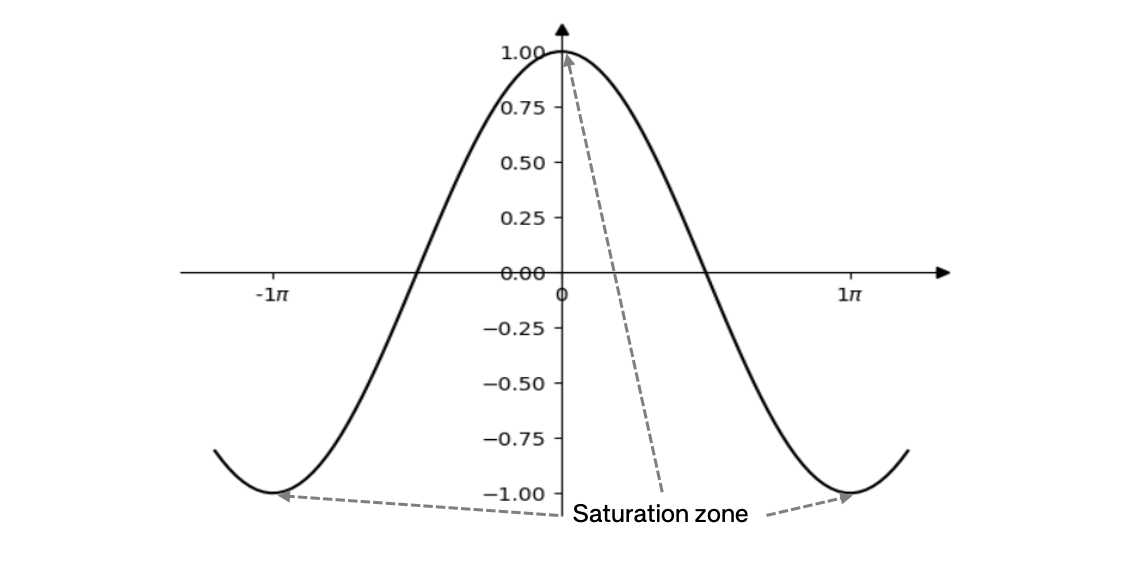}
\caption{Cosine function's saturation zones exhibit near-zero gradients, which makes it difficult for the model to learn during  backpropagation.  } \label{cos}
\end{figure}
\paragraph{\textbf{ Universal AnglE Embedding (UAE) }}  Similar to GTE and BGE, UAE also uses the pre-trained BERT model (uncased BERT base model with 110M parameters) as the backbone model (note: UAE\_large\_V1 uses roberta-large \cite{roberta} as the default backbone). As many existing text embedding works employed the cosine function in their training objective to measure the pairwise semantic similarity, the authors from \cite{li2023angle} point out that there is the gradient vanishing issue due to the saturation zones of cosine function (as shown in Figure \ref{cos}), which hinder the ability to learn subtle distinctions between texts in backpropagation. To deal with the problem of vanishing gradients, a novel angle-optimized text embedding model called AnglE is proposed by introducing angle difference optimization in a complex space which substantially improve the text embedding quality in various scenarios. Given input text embedding pair $(E(q), E(d))$, the chunking strategy \cite{sun2019rotate} is used to get their representations in the complex space $(\mathbf{z}, \mathbf{w})$, followed by the angle difference  $\Delta \theta_{qd}$ between $\mathbf{z}$ and $\mathbf{w}$. Then the angle loss can be defined as:

\begin{equation}
\label{angle}
    \mathcal{L}_{angle} = log\left[1 + \sum_{sim(E(i), E(j)) > sim(E(m), E(n)) } e^ \frac{\theta_{ij} - \theta_{mn}}{\tau}   \right]
\end{equation}

where $sim(E(i),E(j))$  is the similarity between the embedding of $i$ and the embedding of $j$. 
The authors also propose LLM-supervised learning (use LLMs as data annotators to label the pseudo-supervised data) to effectively deal with the  domain-supervised data scarcity problem \cite{li2023angle}. 
Some other details about UAE include:
\begin{itemize}
    \item Training data: the NLI datasets MNLI \cite{mnliwilliams2017broad} and SNLI \cite{snlibowman2015large}, and/or LLM supervised data
    \item Loss function:  AnglE loss:  the combination of cosine objective, in-batch negative objective and angle objective.
    \item Negative sampling: in-batch negative samples and/or hard negatives
    \item Model size: 
    \begin{itemize}
        \item AnglE-BERT: 110M (backbone: uncased BERT),
      \item  UAE\_Large\_V1: 355M (backbone: roberta-large), 
    \end{itemize}
      
\end{itemize}

\paragraph{\textbf{ Matryoshka Representation Learning (MRL) \cite{mrlkusupati2022matryoshka}}} 

Deploying deep representation or text embedding involves two steps: a constant forward pass to compute the representation, and its use for downstream applications \cite{sato2021vertex,varma2019extreme}. The computation costs for the second step rise with the embedding dimensionality, data size, and label space, which can exceed the feature computation cost for large scale systems \cite{dean2009challenges,sun2017revisiting}. The rigid nature of these representations requires high-dimensional embedding vectors for different tasks, even though varying resource and accuracy constraints call for flexibility \cite{mrlkusupati2022matryoshka}.
Given that we can't predict the computational and statistical demands for each subsequent task, fixed-capacity representations/embeddings may not always be suitable and could either exceed or fall short of the task's requirements. Could we create an adaptable representation that can adjust to a variety of downstream tasks with fluctuating computational resources?

MRL introduces a novel method for learning representations of data through a nested structure to induce flexibility in the learned representation, similar to Russian Matryoshka dolls, which encodes information at different granularities and allows a single embedding to adapt to the computational constraints of downstream tasks \cite{mrlkusupati2022matryoshka}. The representation/embedding $z$ is a $d$ dimensional vector, $M = [m_1, m_2, ... d] $ are the chosen dimensions which define different representation sizes. MRL makes each of the first $m$ dimensions $z_{1:m}$ to be independently capable of being a general purpose representation of the data point $x$. Given a labelled dataset $D = \{(x_1, y_1), (x_2, y_2), ... (x_N, y_N)\}$ where $N$ is the datasize and $y_i$ is the label of data $x_i$,  MRL uses standard empirical risk minimization to optimize multi-class classification loss for each nested dimension $m \in M$ using a separate linear classifier, parameterized by $\mathbf{W}^{(m)}$: 
\begin{equation}
\label{mrle}
    \underset{\{\mathbf{W}^{(m)}  \}_{m \in M} , \theta_F}{\min}  \frac{1}{N} \sum_{i \in N} \sum_{m \in M} c_m \cdot \mathcal{L}(\mathbf{W}^{(m)} \cdot F(x_i, \theta_F)_{1:m} ; y_i)
\end{equation}

where $\mathcal{L}$ is the multi-class softmax cross-entropy loss function, $F(\cdot ; \theta_F)$ is the deep neural network to get the representation/embedding $z$, $c_m$ is the importance scales. The authors also show that MRL extends seamlessly to web-scale datasets across vision, language, and vision + language. The experimental results show that MRL can be effectively used for large-scale adaptive classification and retrieval, providing similar accuracy to fixed-feature baseline with a significantly smaller representation size, and offering a more cost-effective and faster adaptive shortlisting and re-ranking system \cite{mrlkusupati2022matryoshka}.

\paragraph{\textbf{ 2D Matryoshka Sentence Embeddings (2dMSE)}} 

Despite MRL's enhanced efficiency, it still necessitates going through all transformer layers before obtaining the transformer based text embedding, leading to significant compute and memory consumption. This raises questions about the impact of the fixed number of transformer layers on representation quality and the feasibility of using intermediate layers for sentence representation. With the aim to enhance the flexibility and scalability of the original MRL's sentence embedding learning,  two-dimensional Matryoshka Sentence Embedding (2DMSE) is proposed in \cite{2dmrlli20242d}.  
2DMSE uses $BERT_{base}$ as backbone to encode text data $x$:
\begin{equation}
\label{2d1}
    \mathbf{X}_l^m = BERT_{1:l} ^{cls}(x)_{1:m} \in R^m
\end{equation}

where $cls$ means the pooling strategy using “CLS” embeddings as the sentence embeddings; $l \in [1, L]$
 denotes the $l$-th layer of the L-layer transformer backbone; $m \in M = [m_1, m_2, ... d] $ (same as MRL) represents the first $m$ dimensions in the $d$-dimensional embeddings.  $l$ allows 2DMSE scaling the encoder model in the dimension in terms of the number of layers while $m$ allows 2DMSE scaling the encoder model in the dimension in terms of the embedding size. To ensure the quality of embeddings, full-capacity embeddings from the last attention layer $\mathbf{X}_L^d$ are trained consistently with the following objective:
 
\begin{equation}
\label{2d2}
   \mathcal{L}_L^d =loss(\mathbf{X}_L^d ; A)
\end{equation}

The auxiliary information A is utilized for loss computation, typically indicating positive or negative samples or providing ranking details \cite{2dmrlli20242d}. During the same training step, a shallower Transformer layer $l$ is randomly chosen following a uniform distribution $l \sim U(1, L-1)$, and its complete embedding vector is directly utilized for representation learning:
\begin{equation}
\label{2d3}
   \mathcal{L}_l^d =loss(\mathbf{X}_l^d ; A)
\end{equation}

2DMSE also uses MRL for nested low-dimensional vectors at both the last layer $\mathbf{X}_L$:

\begin{equation}
\label{2d4}
   \mathcal{L}_L^m =loss(\mathbf{X}_L^m ; A)
\end{equation}

and the sampled layer  $\mathbf{X}_l$:

\begin{equation}
\label{2d5}
   \mathcal{L}_l^m =loss(\mathbf{X}_l^m ; A)
\end{equation}
where $m$ is the MRL embedding size. 

The next step is to improve the shallow layer’s performance by aligning its embeddings to the last layer's:
\begin{equation}
\label{2d6}
   \mathcal{L}_{align} = KLDiv(\mathcal{L}_l^d, \mathcal{L}_L^d) + KLDiv(\mathcal{L}_l^m, \mathcal{L}_L^m) 
\end{equation}

where KLDiv(,) denotes the Kullback-Leibler divergence. The weighted sum of [$\mathcal{L}_L^d$, $\mathcal{L}_l^d$, $\mathcal{L}_L^m$, $\mathcal{L}_l^m$, $\mathcal{L}_{align}$] is used as the final objective. 
 
Based on MRL and 2DMSE, several well performing text embeddings including mxbai-embed-large-v1 (335M) and mxbai-embed-2d-large-v1 (335M) are released in \cite{emb2024mxbai}. However, the training details of these models are not documented. 

\paragraph{\textbf{ Summary}} 
In this section, the MTEB top performing universal text embedding models with the focus on proposing new loss functions are introduced. Apart from proposing variants on the classic InfoNCE loss, UAE introduces AnglE loss by introducing angle optimization in a complex space to deal with the vanishing gradients problem from cosine function's saturation zone. Another line of research focuses on  adaptable representations that can
adjust to a variety of downstream tasks with fluctuating computational resources, where MRL proposes new loss function to make each of the first $m$ dimensions of the text embedding to be independently capable of being a general purpose representation and 2dMSE proposes new loss function based on MRL to make each of the first $m$ dimensions of each layer of the transformer of the text embedding to be independently capable of being a general purpose representation.

\section{LLMs focused universal text embeddings}

\begin{table*}[t!]
\caption{The comparison among LLM focused universal text embedding models. Some methods test multiple backbone models, only the best performing ones are listed. LLM gen data indicates whether synthetic data generated by LLMs are used to train the model. The sign - means no information available. }\label{t2}
\centering
\small
\begin{tabular}{p{1.5cm}  p{1.4cm}   p{3.5cm}   p{2.2cm}  p{3.1cm} p{1.cm}  }

\\ \hline
Models & Backbone & Key contributions & Fine-tune strategy & Fine-tune efficiency  & LLM gen data  \\
\hline
E5-mistral-7b-instruct  & Mistral-7b & Fine-tune decoder only LLMs with a mix of real and synthetic data generated by LLMs & LoRA with rank 16 (42M parameters); Batch size: 2048 & 576 GPU hours on V100 GPU (18 hours on 32 V100 GPUs ) & Yes  \\ \hline
SFR-Embedding-Mistral & Mistral-7b & Multi-task finetuning over E5-mistral-7b-instruct with improved hard negatives & LoRA with rank 8 (21M parameters); Batch size: 2048 & 120 GPU hours on A100 GPU (15 hours on 8 A100 GPUs) & Yes \\ \hline
Echo-mistral  & Mistral-7b & Use bidirectional attention: repeat the input twice and extract embeddings from the second occurrence. & LoRA with rank 16 (42M parameters); Batch size: 2048 & 192 GPU hours on A100 GPU (two days on 4 A100 GPUs) & No  \\ \hline
LLM2Vec  & Llama-3    Mistral-7b & Enabling bidirectional attention + Masked next token prediction + Unsupervised contrastive learning & LoRA with rank 16  & - & No  \\ \hline
GRIT & Mistral-7b Mistral-8x7b & Unify generative and embedding tasks by distinguishing between them through instructions & Batch size: 2048 for embedding data; 256 for generative data & 7B model: 3072 GPU hours on A100 80GB GPU; 8X7B model: 20,480 GPU hours on H100 80GB GPU & Yes \\ \hline
Gecko  & gtr-t5-xl (1.2B, encoder from T5-3B model) & Use LLMs to generate Few-shot Prompted Retrieval dataset (FRet) to improve text embedding models & - & - & Yes  \\ \hline
gte-Qwen1.5-7B-instruct & Qwen1.5-7B & Use bidirectional attention along with a vast, multilingual, diverse text corpus  & - & - & -  \\ \hline

\end{tabular}
\end{table*}

LLMs are are extensively pre-trained on diverse large quantity of web-scale data, which can be  used to improve the universal text embeddings in two different ways as summarized in Table \ref{t2}. Firstly, LLMs can be used to generate high quality multilingual multi-task synthetic data as demonstrated by researchers from Microsoft and Google \cite{e5mistralwang2023improving,e5instructwang2024multilingual,lee2024gecko}.  Secondly, LLMs can be used as backbone for text embeddings as they do not need the contrastive pre-training step used in existing state of the art text embedding models. For example, E5-mistral-7b-instruct perform multi-task fine-tuning on Mistral 7b model which is one of the best performing method on MTEB. Echo-mistral \cite{echoispringer2024repetition}, LLM2Vec \cite{llm2vec}, gte-Qwen1.5-7B-instruct \cite{gteli2023towards} and GRIT \cite{gritmuennighoff2024generative} propose various different solutions so that decoder only LLMs with casual attention can generate high quality text embeddings using bidirectional attention. More details about how different universal text embeddings leverage LLMs to improve their universality can be found below.

\paragraph{\textbf{ E5-mistral-7b-instruct }} 

E5-mistral-7b-instruct is one of the best performing text embeddings on the MTEB benchmark, which is also a representative text embedding model leveraging LLMs. Firstly, proprietary LLMs including GPT-35-Turbo and GPT-4 are used to generate  synthetic data covering a diverse range of text embedding tasks in 93 languages (among which 25\% are generated by GPT-35-Turbo and others are generated by GPT-4) \cite{e5mistralwang2023improving}. In terms of the quality generated data, the authors find that the overall quality is acceptable despite a portion of GPT-35-Turbo outputs do not follow the instructions in the prompt templates strictly. Secondly, pre-trained open source LLM Mistral-7b checkpoint \cite{jiang2023mistral} is selected to be fine-tuned on a mixture of synthetic and labeled data (collection of 13 public datasets) with around 1.8M examples after sampling.  One advantage of using LLMs such as Mistral \cite{jiang2023mistral} for text embedding is that they are extensively pre-trained on web-scale data already, which does not need the contrastive pre-training step used in existing state of the art text embedding models. 
Given a pre-trained LLM, an [EOS] token is appended to the end of the query and document. The last layer [EOS] vector is used as the text embeddings.
To help the model accommodate different tasks, instruction templates (which are used by all LLMs focused universal text embeddings described in this section as well as some of the previously mentioned universal text embeddings such as BGE \cite{bgexiao2023c} ) are applied to the original query $q^+$ to generate a new one $q_{inst}^+$ given a relevant query-document pair $(q^+, d^+)$:
\begin{equation}
\label{inst}
    q_{inst}^+ = Instruct: \{task\_definition\} \;  \setminus n  \; \;  Query: \{q^+\} 
\end{equation}
where “{task\_definition}” is a placeholder for a one-sentence description of the embedding task added only to the query side but not to the document side \cite{e5mistralwang2023improving}. 
Some other details about E5-mistral-7b-instruct include:
\begin{itemize}
    \item Fine-tuning data: generated synthetic data, ELI5 \cite{fan2019eli5} (sample ratio 0.1), HotpotQA \cite{yang2018hotpotqa}, FEVER \cite{thorne2018fever}, MIRACL \cite{zhang2023miracl}, MS-MARCO passage ranking (sample ratio 0.5) and document ranking (sample ratio 0.2) \cite{bajaj2016ms}, NQ \cite{nq1karpukhin2020dense}, NLI \cite{snlibowman2015large}, SQuAD \cite{nq1karpukhin2020dense}, TriviaQA \cite{nq1karpukhin2020dense}, Quora Duplicate Questions \cite{iyer2017quora} (sample ratio 0.1), Mr-TyDi \cite{zhang2021mr}, DuReader \cite{qiu2022dureader_retrieval}, and T2Ranking \cite{xie2023t2ranking} (sample ratio 0.5) datasets.
    \item Loss function: standard InfoNCE loss as in Equation \ref{infonce}
    \item Negative sampling:  in-batch negative samples and hard negatives (for the datasets without hard negatives,  mE5base \cite{e5instructwang2024multilingual} is used to to mine top 100 hard negatives).
    \item Model size: 7B (42M trainable parameters using Low-rank adaptation (LoRA) \cite{hu2021lora})
      
\end{itemize}

The experimental results from \cite{e5mistralwang2023improving} shows that even with only synthetic data, the performance of E5-mistral-7b-instruct on MTEB English benchmark is still very competitive. E5-mistral-7b-instruct also has the multilingual capabilities with good performances over high-resource languages. Furthermore, the authors discovered that the method of incorporating instructions has a considerable impact on the performance. Their hypothesis is that the model is better informed about the embedding task at hand through natural language instructions, thereby allowing the model to produce more distinctive embeddings \cite{e5mistralwang2023improving}.

 \paragraph{\textbf{ SFR-Embedding-Mistral }} Built on top of the E5-mistral-7b-instruct,  SFR-Embedding-Mistral is also one of the top-ranking universal text embeddings on the MTEB English benchmark with 0.93\% performance increase compared to  E5-mistral-7b-instruct. The authors summarized their main takeaways in \cite{SFRAIResearch2024} (the detailed report is not released) as:  
\begin{itemize}
    \item The retrieval performance of text embeddings significantly improves when integrated with clustering tasks and further enhanced through multi-task knowledge transfer. 
    \item Task-homogeneous batching, a method that forms batches from a single task, improves the performance of text embedding by making in-batch negatives more challenging.
    \item Improving the construction of hard negatives enhances the model's capacity to accurately identify misleading documents.
\end{itemize}

To be noted that, the following multi-task datasets are used by SFR-Embedding-Mistral  to fine-tune the E5-mistral-7b-instruct model, including 
\begin{itemize}
    \item Retrieval tasks: MS-MARCO, NQ, FiQA, SciFact, NFCorpus, DBPedia, FEVER, HotpotQA, Quora and NLI.
    \item Clustering tasks: arXiv, bioRxiv,  medRxiv, applying filters to exclude development and testing sets in the MTEB clustering framework.
    \item Classification tasks: AmazonReview, Emotion, MTOPIntent, ToxicConversation , and TweetSentiment.
    \item Semantic Textual Similarity (STS) tasks: STS12, STS22 , and STSBenchmark
    \item Reranking tasks: SciDocs and StackOverFlowDupQuestions.
\end{itemize}
Among the selected training datasets, most are from the MTEB benchmark. Even the development and testing sets are excluded, it might have an unfair advantage comparing to other text embedding methods that do not use the MTEB training data.

\paragraph{\textbf{ Echo-mistral}} 

Even though constructing text embeddings from autoregressive pretrained LLMs seems promising, the authors from \cite{echoispringer2024repetition} identified a striking failure mode of autoregressive language models trained on the next-token objective: the contextualized token embeddings, represented by the vector of last-hidden-layer activations at a specific input token's position, lack information from tokens appearing later in the sentence because of the causal attention mask. Given the following example provided in \cite{echoispringer2024repetition}:
\begin{itemize}
    \item q: \textcolor{blue}{[She loves summer]} \textcolor{red}{[but dislikes the heat]}
    \item $d^-$: \textcolor{blue}{[She loves summer]} \textcolor{red}{[for the warm evenings]}
    \item $d^+$: \textcolor{blue}{[She loves summer]} \textcolor{red}{[but not the temp]}
\end{itemize}
In this example, the classical LLMs based contextualized embeddings of the first half of $d^-$ and $d^+$ are both similar to q because they do not attend to the second half of the sentence, which leads to the  overestimation of the similarity between q and  $d^-$ by any pooling strategy that uses information from the first half \cite{echoispringer2024repetition}.

To mitigate this striking failure mode and take advantage of the bidirectional context information, a simple fix is proposed by presenting the input sentence twice to LLMs.  The final contextualized embeddings can then be extracted from the second occurrence of the sentence.  LLMs are instructed to undertake  basic task such as rewriting or repeating in order to prompt the second occurrence to effectively "encode" information from the first \cite{echoispringer2024repetition}. 
Despite twice the computational cost of classical embeddings, experimental results show that Echo embeddings can improve the LLM based text embedding quality significantly under both zero-shot setting and fine-tuning setting. 
Some other details about Echo-mistral (echo-mistral-7b-instruct-last) include:
\begin{itemize}
    \item Fine-tuning data: same as E5-mistral-7b-instruct \cite{e5mistralwang2023improving} without synthetic data
    \item Loss function: standard InfoNCE loss as in Equation \ref{infonce}
    \item Negative sampling:  in-batch negative samples and mined hard negatives 
    \item Model size: 7B     
\end{itemize}

\paragraph{\textbf{ LLM2Vec }} 
Similar to the idea of Echo-mistral, the authors of \cite{llm2vec} believe that the slow adoption of decoder-only LLMs in text embedding tasks is partly due to their causal attention mechanism, which restricts their ability to create bidirectional contextualized representations from encompassing information from the whole input sequence (a necessary trade-off for generative capabilities). Improving the architectural flaw of decoder-only LLMs for text embedding tasks is highly desirable because: 1. decoder-only LLMs are much more sample-efficient than encoder-only models \cite{clark2020electra}; 2. LLMs are supported by a robust ecosystem, including comprehensive tools and well tested pre-training techniques, leading to their continuous enhancement by the community; 3. the good instruction following ability of LLMs \cite{wang2022super,ouyang2022training} makes them ideal for creating universal text embedding models that can handle a wide range of tasks using instructions.

To improve the text embeddings from decoder-only LLMs, LLM2Vec proposes a simple unsupervised approach that can transform any decoder-only LLM into a strong text encoder in three simple steps: 1. enabling bidirectional attention by replacing the causal attention mask of decoder-only LLMs with an all-ones matrix; 2. Masked Next Token Prediction (MNTP): combining next token prediction with masked language modeling \cite{devlin2018bert} to make the model aware of its bidirectional attention; and 3. unsupervised contrastive learning for better sequence representations: the model processes an input sequence twice with independently sampled dropout masks to generate two distinct representations, and is trained to increase the similarity between these representations while decreasing similarity with other sequence representations in the batch \cite{llm2vec}. Their empirical results show that LLMs can be efficiently converted into universal text embeddings without requiring costly adaptation or synthetic GPT-4 generated data.  Some other details about LLM2Vec include:

\begin{itemize}
    \item Unsupervised training data: English Wikipedia
    \item Supervised contrastive learning data: adaptations of E5 \cite{e5mistralwang2023improving}: the public portion of the E5 dataset \cite{e5mistralwang2023improving} curated by \cite{e5instructwang2024multilingual}
    \item Loss function: Contrastive loss, masked next token prediction loss
    \item Negative sampling:  in-batch negatives and hard negatives 
    \item Model size: the best performing model of LLM2Vec on MTEB is LLM2Vec-Mistral7B-Ins-v2-sup (backbone: Mistral 7B): 7B
\end{itemize}

\paragraph{\textbf{ Generative Representational Instruction Tuning (GRIT)}} 

Similar to the idea from Echo-mistral and LLM2Vec, the authors in \cite{gritmuennighoff2024generative} also highlight the importance of bidirectional attention for general purpose universal text embeddings. However, GRIT takes the general purpose model to the next level by training a large language model   to handle both generative and embedding tasks (all text-based language problems) distinguished through instructions. 

Both representational instruction tuning \cite{su2022one,e5mistralwang2023improving,asai2022task}  and generative instruction tuning \cite{muennighoff2022crosslingual,sanh2021multitask,wei2021finetuned} are combined into one unified model by GRIT. Firstly, GRIT uses bidirectional attention with mean pooling over the final hidden state to get the text embedding. Contrastive objective with in-batch negatives are used to finetune a pretrained large language model following prior works \cite{chen2020simple,nligao2021simcse}.  The average of the final hidden states of only the input sample is calculated during mean pooling, while disregarding the instruction and format tokens. Nonetheless, these tokens continue to impact the final representation via the self-attention mechanism \cite{vaswani2017attention}.
Secondly,  the language modeling objective of next token prediction \cite{gpt1radford2018improving} is used to compute the loss on generative data, where a language modeling head on top of the hidden states predicts the next tokens \cite{gritmuennighoff2024generative}. Finally, the representational and generative objectives are summed with optional loss weights. Furthermore, sliding window attention \cite{child2019generating,beltagy2020longformer}  is used by GRIT to handle generative and embedding inputs of arbitrary length.

The primary drawback of GRIT is its increased computational demand (as shown in Table \ref{t2}), resulting from the need to train with two objective functions. GRITLM 7B is fine-tuned from Mistral 7B \cite{jiang2023mistral} and GRITLM 8x7B \cite{jiang2024mixtral} is fine-tuned from Mistral 8x7B. Both models have top performance on MTEB English benchmark.  GRITLM 7B has better performance than GRITLM 8X7B on embedding tasks, while GRITLM 8X7B is significantly better than GRITLM 7B on generative tasks. The authors provide some hypothesis on the reason why GRIT works on both embedding and generative tasks: 1. Generative language modeling and text embeddings are interconnected, requiring deep understanding of natural language, but expressed differently. 2. The unified model may contain parameters acting as a switch for either embedding tasks or generative tasks \cite{gritmuennighoff2024generative}.
Some other details about GRIT include:
\begin{itemize}
    \item Fine-tuning data: adaptations of E5 \cite{e5mistralwang2023improving}: adding S2ORC \cite{lo2019s2orc} to increase its scientific data (“E5S”); adaptations of Tülu 2 data \cite{ivison2023camels}: filtering out their custom prompts that contain answers related to the origin of their model.
    \item Loss function: Contrastive loss with next token prediction loss
    \item Negative sampling:  in-batch negative samples and hard negatives
    \item Model size: 
    \begin{itemize}
        \item GRITLM 7B: 7B 
        \item GRITLM 8X7B: 47B
    \end{itemize}     
\end{itemize}

\paragraph{\textbf{ Gecko }} 
LLM based text embeddings have several disadvantages including high computational cost, longer response time and high embedding dimensions (which makes the downstream tasks training also computationally expensive). A recent paper named Versatile Text Embeddings Distilled from Large Language Models \cite{lee2024gecko} tries to  mitigate these limitations using   knowledge distillation from LLMs with synthetic data generation and refinement, where queries are generated from LLMs given the contexts, and their positive and negative passages are mined and refined by LLMs.  

The main contribution of \cite{lee2024gecko} is designing the two-stage approach that uses LLMs to generate Few-shot Prompted Retrieval dataset (FRet). The first stage is LLM-based Diverse Query Generation: unlike \cite{e5mistralwang2023improving}, FRet uses LLM to analyze a selected web passage and produce both a description of the task $t$ and a pertinent query $q$ related to the task: 
 \begin{equation}
\label{gecko1}
   LLM (P_{QG}, p_{seed}) \xrightarrow{} (t,q)
\end{equation}
 where $p_{seed}$ is a passage drawn randomly from the web corpus $\mathcal{C}$ and $P_{QG}$ is a fixed few-shot prompt that is identical for every example.  By drawing from a variety of free-form task descriptions,  LLM is guided to generate a diverse set of queries. These pairs are subsequently utilized to train the embedding models, instructing the models to link a query and its associated instructions with the target passage \cite{lee2024gecko}. To further encourage the diversity of generated task descriptions and queries,  many diverse task descriptions are added in the prompt. 

 The second stage of FRet is LLM-based positive and negative mining. Unlike previous works \cite{jeronymo2023inpars,dai2022promptagator} assuming that the query $q$ generated from a given passage $p_{seed}$ forms a good training pair $(q, p_{seed})$, the authors hypothesize that there could be a better positive target passage for $q$ than $p_{seed}$ somewhere in the corpus of web passages as $p_{seed}$ is not  guaranteed to maximize $P(p|q, t)$ over all the passages in the corpus \cite{llm2vec}. To mine better positives for the generated query, they train an initial embedding model using passage and generated query pairs $(q, p_{seed})$ with in-batch negatives.  The trained embedding model is used to retrieve top $K$ neighbors $P = \{ p^{(1)}, ..., p^{(K)} \}$ from the corpus given a generated query $q$. These retrieved passages are ranked by the LLM with two few-shot prompted LLM ranking functions: 
 \begin{itemize}
     \item  Query Likelihood (QL)  \cite{sachan2022improving}: $QL(q,p) = LLM (q|p, \mathcal{P}_{QL})$, where $ \mathcal{P}_{QL}$ is a prompt containing an instruction for judging query likelihood and several few-shot examples of relevant query and passage pairs \cite{drozdov2023parade}.
     \item Relevance Classification (RC) \cite{zhuang2023beyond}: $RC(q,p) = LLM (label|p, \mathcal{P}_{RC})$, where $\mathcal{P}_{RC}$ is a prompt with few-shot examples for grading the relevance of each query-passage pair.
 \end{itemize}

The final ranking function $R(q, p)$ is obtained by combining the rankings from QL and RC with the standard Reciprocal Rank Fusion (RRF) approach \cite{cormack2009reciprocal}. The top ranked passage is then selected as the new positive passage $p^+$ given the generated query ($p^+ \neq p_{seed}$ happens for around 15\% cases in their dataset). In terms of negative passage selection, they propose two methods: 1.  a random nearest neighbor passage that is different from the original passage; 2. the k-th passage in the ranking. The generated  FRet dataset has in total 6.6M examples, each containing a task, a query, a positive passage, and a negative passage \cite{lee2024gecko}. The authors propose a new embedding model Gecko based on a 1.2B parameter pre-trained transformer language model and fine-tuned on the generated FRet dataset, which is one of the top performing text embeddings on the English MTEB benchmark with small embedding dimensions (256 and 768). Some other details about Gecko include:
\begin{itemize}
    \item Pre-training data: large-scale community QA dataset \cite{gtrni2021large} with title-body text pairs from the Web.
    \item Unified fine-tuning data: FRet (6.6M) along with the following academic training datasets: Natural Questions \cite{nq2kwiatkowski2019natural}, HotpotQA \cite{yang2018hotpotqa}, FEVER \cite{thorne2018fever}, MedMCQA \cite{pal2022medmcqa}, SNLI \cite{snlibowman2015large}, MNLI \cite{mnliwilliams2017broad}, and several classification datasets from Huggingface. For the multilingual model,  training sets from MIRACL \cite{zhang2023miracl} is added. 
    \item Loss function: 
    \begin{itemize}
        \item Pre-training: the contrastive loss
        \item Fine-tuning:    in-batch cross-entropy loss
    \end{itemize}  
    \item Negative sampling: 
    \begin{itemize}
        \item Pre-training: in-batch negative samples
        \item Fine-tuning: in-batch negative samples,  hard negatives and same-tower negatives (other queries in the batch) \cite{moiseev2023samtone}
    \end{itemize}
    \item Model size (google-gecko-preview-0409): 1.2B (backbone: gtr-t5-xl \cite{gtrni2021large})
      
\end{itemize}

\paragraph{\textbf{ gte-Qwen1.5-7B-instruct}}  
The authors from GTE \cite{gteli2023towards} also proposed  their LLMs focused universal text embedding model based on Qwen1.5-7B large language model \cite{qwen}, which is one of the the top-ranking embedding models on both MTEB English and Chinese benchmarks. While the full details of fine-tuning are not disclosed, the authors summarized their key contributions as \footnote{https://huggingface.co/Alibaba-NLP/gte-Qwen1.5-7B-instruct}: 1. the use of bidirectional attention mechanisms to enhance contextual understanding;  2. the use of instruction tuning; 3. the use of a large, multilingual text corpus that covers various domains and scenarios.

\paragraph{\textbf{ Summary}} 
In this section, the universal text embeddings leveraging LLMs 
(which make up the majority of the top 10 best performing models on the MTEB English benchmark)  are  introduced. Most of these models share the finding that  LLMs acquire good text representations through comprehensive auto-regressive pre-training, requiring only minimal fine-tuning to become effective universal text embedding models. E5-mistral-7b-instruct from Microsoft and Gecko from Google DeepMind demonstrate two different ways to generate synthetic data from LLMs in order to improve universal text embeddings, while Echo-mistral \cite{echoispringer2024repetition} and LLM2Vec \cite{llm2vec} show that good universal text embeddings can be achieved with the focus on enabling the bidirectional attentions for decoder only LLMs without using synthetic data. LoRA is widely used for the fine-tuning of LLMs based universal text embeddings, where LoRA ranks are found not affecting the overall performance substantially in \cite{e5mistralwang2023improving}. Instructions are used by all LLM focused text embedding models introduced in this section. One of the main reasons is the good instruction following ability of LLMs which makes them ideal for creating universal text embedding models that can handle a wide range of tasks using instructions. 
From Table \ref{t2}, it can be told that Mistral-7B model is the most popular backbone model for LLMs focused text embeddings. One of the reasons is that enabling bidirectional attention (even without any training) works well for Mistral-7B. The authors from \cite{llm2vec} speculate that Mistral models may be pre-trained with some form bidirectional attention. 
On the other hand, the full evaluation on MTEB of LLM based universal text embedding models is reported to be computationally expensive: it takes about 3 days on 8 V100 GPUs for E5-mistral-7b-instruct and 40 hours on 8x A100 GPUs for LLM2Vec with Mistral-7B as backbone.

\section{Analysis on performances and limitations}
\subsection{Overall performance on MTEB English benchmark}

\begin{table*}
\caption{The top 25 best performing text embeddings methods on MTEB English benchmark. Model size is measured in Million Parameters, \#Memory is Memory Usage measured in (GB, fp32), \#Embedding is the Embedding dimension. Results can be found from HuggingFace website: https://huggingface.co/spaces/mteb/leaderboard.} \label{tab1}
 \centering
 \small
\begin{tabular}{p{3.8cm}  p{1.cm}   p{1.2cm}   p{1.3cm}  p{2.cm}   p{1.2cm}   p{1.2cm}  }
\\ \hline
model\_names & rank & Model Size & \#Memory & \#Embedding & Max Tokens & Average\\
 \hline
SFR-Embedding-Mistral & 1 & 7111 & 26.49 & 4096 & 32768 & 67.56 \\ \hline
voyage-lite-02-instruct & 2 & 1220 & 4.54 & 1024 & 4000 & 67.13 \\ \hline
GritLM-7B & 3 & 7242 & 26.98 & 4096 & 32768 & 66.76 \\ \hline
e5-mistral-7b-instruct & 4 & 7111 & 26.49 & 4096 & 32768 & 66.63 \\ \hline
google-gecko-preview-0409 & 5 & 1200 & 4.47 & 768 & 2048 & 66.31 \\ \hline
GritLM-8x7B & 6 & 46703 & 173.98 & 4096 & 32768 & 65.66 \\ \hline
LLM2Vec-Mistral7B-Ins-v2-sup & 7 & - & - & - & - & 64.80 \\ \hline
echo-mistral-7b-instruct-last & 8 & 7111 & 26.49 & 4096 & 32768 & 64.68 \\ \hline
mxbai-embed-large-v1 & 9 & 335 & 1.25 & 1024 & 512 & 64.68 \\ \hline
UAE-Large-V1 & 10 & 335 & 1.25 & 1024 & 512 & 64.64 \\ \hline
text-embedding-3-large & 11 & - & - & 3072 & 8191 & 64.59 \\ \hline
voyage-lite-01-instruct & 12 & - & - & 1024 & 4000 & 64.49 \\ \hline
Cohere-embed-english-v3.0 & 13 & - & - & 1024 & 512 & 64.47 \\ \hline
multilingual-e5-large-instruct & 14 & 560 & 2.09 & 1024 & 514 & 64.41 \\ \hline
google-gecko-256-preview-0409 & 15 & 1200 & 4.47 & 256 & 2048 & 64.37 \\ \hline
GIST-large-Embedding-v0 & 16 & 335 & 1.25 & 1024 & 512 & 64.34 \\ \hline
bge-large-en-v1.5 & 17 & 335 & 1.25 & 1024 & 512 & 64.23 \\ \hline
LLM2Vec-Llama2-7b-sup & 18 & - & - & - & - & 64.14 \\ \hline
Cohere-embed-multilingual-v3.0 & 19 & - & - & 1024 & 512 & 64.01 \\ \hline
GIST-Embedding-v0 & 20 & 109 & 0.41 & 768 & 512 & 63.71 \\ \hline
bge-base-en-v1.5 & 21 & 109 & 0.41 & 768 & 512 & 63.55 \\ \hline
ember-v1 & 22 & 335 & 1.25 & 1024 & 512 & 63.54 \\ \hline
sf\_model\_e5 & 23 & 335 & 1.25 & 1024 & 512 & 63.34 \\ \hline
mxbai-embed-2d-large-v1 & 24 & 335 & 1.25 & 1024 & 512 & 63.25 \\ \hline
gte-large & 25 & 335 & 1.25 & 1024 & 512 & 63.13 \\
 \hline
\end{tabular}
\end{table*}

Due to the differences in training data, back-bone model, loss function, training strategy, negative-sampling strategy, embedding dimension and so on, it is difficult to have a fair comparison among different text embedding models. But we can still get some insights from the overall performance comparison.   
The overall performance of the top 25 best text embeddings methods on MTEB English benchmark are shown in Table \ref{tab1}, where the Model size is measured in Million Parameters, \#Memory is Memory Usage measured in (GB, fp32), \#Embedding is the Embedding dimension. It can be seen that some of the top performing text embeddings are not introduced in this review including voyage-lite-02-instruct, voyage-lite-01-instruct, text-embedding-3-large, Cohere-embed-english-v3, Cohere-embed-multilingual-v3, ember-v1, sf\_model\_e5, etc.  The main reason is that these models do not disclose any detailed documentation.  

For the models with documentations available, it can be seen that SFR-Embedding-Mistral has the best performance\footnote{Note that this might change due to new models added to the MTEB benchmark.}, with the average performance over 56 MTEB datasets of 67.6\%. SFR-Embedding-Mistral increases the performance over e5-mistral-7b-instruct by 0.93\% by fine-tuning on top of e5-mistral-7b-instruct using more datasets including MTEB training data.  GritLM-7B is ranked the 3rd place, outperforming  GritLM-8x7B by 1.1\%, even though GritLM-8x7B has much more parameters (46.7B parameters) than GritLM-7B (7.2B parameters).  To be noted that, GritLM-7B and GritLM-8x7B has unified both text embedding and text generation in the same model, which is different from other text embedding models. Among the top 5 performing text embeddings,  google-gecko-preview-0409 and voyage-lite-02-instruct have the smallest parameters (around 1.2B), while google-gecko-preview-0409 has the smallest embedding dimension which is favored by downstream tasks. LLM2Vec-Mistral7B-Ins-v2-sup and echo-mistral-7b-instruct-lasttoken both use Mistral 7B as backbone and both focus on making decoder only LLMs use bidirectional attention to get better text embeddings. Even though their performances are similar, LLM2Vec-Mistral7B-Ins-v2-sup has the advantage of being more computational efficient. 

Starting from mxbai-embed-large-v1 ranked at the 9th place till gte-large ranked at 25th place, most text embedding models are BERT based with relatively smaller model size compared to LLM based text embeddings. Both mxbai-embed-large-v1 (rank 9) and UAE-Large-V1 (rank 10) propose innovative loss function improvement in the field. GIST-large-Embedding-v0 (rank 16) is built on top of bge-large-en-v1.5 (rank 17) with improvement on in-sample selection of negatives as well as the usage of MTEB training data.   gte-large, bge-base-en-v1.5, bge-large-en-v1.5 and multilingual-e5-large-instruct models show the strong performance of BERT based models with smaller model size and embedding dimensions. Among the top 25 text embeddings, google-gecko-256-preview-0409 has the smallest embedding dimension (256) but still has good performance (rank 15). 

\subsection{The universality analysis}
\begin{table*}
\caption{The improvements over different tasks of the top performing text embeddings compared to the baseline method SimCSE. Each text embedding model's performance is divided by the baseline performance in the table: 1 means the model has the same performance as the baseline, larger than 1 values means the model improves the performance of the baseline, smaller than 1 values means the baseline outperforms the model. Classi is short for Classification task, Pair-C is short for Pair Classification task and Summa is short for Summarization task in this table.}\label{tab2}
 \centering
 \small
\begin{tabular}{p{3.5cm}  p{0.8cm}   p{0.8cm}   p{1.3cm}  p{1.cm}   p{1.3cm}   p{1.1cm}  p{0.8cm}  p{0.9cm}  }
\\ \hline
model\_names & Avg & Classi & Clustering & Pair-C & Reranking & Retrieval & STS & Summa\\
 \hline
SFR-Embedding-Mistral & \textbf{1.3824} & 1.1635 & 1.5456 & \textbf{1.2017} & \textbf{1.2756} & \textbf{2.7039} & 1.0749 & 0.9997 \\ \hline
voyage-lite-02-instruct & 1.3736 & 1.1772 & \textbf{1.5681} & 1.1790 & 1.2251 & 2.5940 & \textbf{1.0843} & 0.9949 \\ \hline
GritLM-7B & 1.3661 & 1.1803 & 1.5139 & 1.1830 & 1.2724 & 2.6311 & 1.0535 & 0.9743 \\ \hline
e5-mistral-7b-instruct & 1.3634 & 1.1656 & 1.5034 & 1.1990 & 1.2665 & 2.6072 & 1.0696 & 1.0074 \\ \hline
google-gecko-preview-0409 & 1.3569 & \textbf{1.2057} & 1.4203 & 1.1891 & 1.2390 & 2.5527 & 1.0752 & 1.0468 \\ \hline
GritLM-8x7B & 1.3436 & 1.1665 & 1.4999 & 1.1532 & 1.2579 & 2.5247 & 1.0523 & 0.9567 \\ \hline
LLM2Vec-Mistral7B-Ins-v2-sup & 1.3260 & 1.1383 & 1.3622 & 1.1942 & 1.2289 & 2.5660 & 1.0628 & 0.9612 \\ \hline
echo-mistral-7b-instruct-lasttoken & 1.3235 & 1.1502 & 1.3856 & 1.1854 & 1.2230 & 2.5445 & 1.0435 & 0.9859 \\ \hline
mxbai-embed-large-v1 & 1.3235 & 1.1236 & 1.3972 & 1.1835 & 1.2644 & 2.4927 & 1.0743 & \textbf{1.0494} \\ \hline
UAE-Large-V1 & 1.3227 & 1.1227 & 1.3978 & 1.1842 & 1.2596 & 2.5050 & 1.0685 & 1.0276 \\ \hline
text-embedding-3-large & 1.3217 & 1.1208 & 1.4660 & 1.1634 & 1.2444 & 2.5408 & 1.0330 & 0.9599 \\ \hline
voyage-lite-01-instruct & 1.3196 & 1.1110 & 1.4179 & 1.1749 & 1.2566 & 2.5472 & 1.0482 & 0.9936 \\ \hline
Cohere-embed-english-v3.0 & 1.3192 & 1.1362 & 1.4188 & 1.1650 & 1.2202 & 2.5206 & 1.0442 & 0.9682 \\ \hline
multilingual-e5-large-instruct & 1.3180 & 1.1521 & 1.4089 & 1.1698 & 1.2322 & 2.4047 & 1.0715 & 0.9750 \\ \hline
google-gecko-256-preview-0409 & 1.3172 & 1.1735 & 1.3482 & 1.1842 & 1.2154 & 2.4033 & 1.0734 & 1.0382 \\ \hline
GIST-large-Embedding-v0 & 1.3166 & 1.1291 & 1.3925 & 1.1767 & 1.2631 & 2.4491 & 1.0691 & 0.9933 \\ \hline
bge-large-en-v1.5 & 1.3143 & 1.1285 & 1.3784 & 1.1824 & 1.2627 & 2.4881 & 1.0504 & 1.0141 \\ \hline
LLM2Vec-Llama2-7b-sup & 1.3125 & 1.1338 & 1.3533 & 1.1948 & 1.2070 & 2.5023 & 1.0583 & 0.9140 \\ \hline
Cohere-embed-multilingual-v3.0 & 1.3098 & 1.1291 & 1.3940 & 1.1692 & 1.2171 & 2.4675 & 1.0509 & 0.9942 \\ \hline
GIST-Embedding-v0 & 1.3037 & 1.1294 & 1.3823 & 1.1716 & 1.2488 & 2.3973 & 1.0555 & 0.9904 \\ \hline
bge-base-en-v1.5 & 1.3004 & 1.1220 & 1.3691 & 1.1747 & 1.2381 & 2.4404 & 1.0415 & 0.9968 \\ \hline
ember-v1 & 1.3002 & 1.1288 & 1.3634 & 1.1858 & 1.2629 & 2.3795 & 1.0533 & 0.9888 \\ \hline
sf\_model\_e5 & 1.2961 & 1.0986 & 1.3943 & 1.1787 & 1.2592 & 2.3740 & 1.0598 & 1.0141 \\ \hline
mxbai-embed-2d-large-v1 & 1.2943 & 1.1013 & 1.3781 & 1.1657 & 1.2398 & 2.3566 & 1.0731 & 1.0122 \\ \hline
gte-large & 1.2918 & 1.0893 & 1.4011 & 1.1536 & 1.2438 & 2.3932 & 1.0535 & 1.0157 \\
 \hline
\end{tabular}
\end{table*}

The pursuit of developing a unified model to address a multitude of downstream tasks has been long-standing \cite{gteli2023towards}. Despite attempting to be general-purpose in previous models such as \cite{cer2018universal,t5raffel2020exploring,st5ni2021sentence}, studies indicate that these embedding models struggle to generalize across tasks and domains \cite{lee2024gecko}. 
In this section, we study whether the MTEB top performing text embeddings are becoming more universal  due to the increasing number and improved quality of diverse text datasets across different tasks \cite{bgexiao2023c,asai2022task}, good quality synthetic data generated by LLMs \cite{lee2024gecko,e5mistralwang2023improving} as well as larger backbones such as LLMs. 

SimCSE (2021) \cite{nligao2021simcse} is selected as the baseline method as it is one of the cornerstone work in text embedding which is cited and adopted by most of the recent works. The improvements over different tasks of the top performing text embeddings compared to the baseline method SimCSE is shown in Table \ref{tab2}. Each text embedding model's performance is divided by the baseline performance in the table: 1 means the model has the same performance as the baseline, larger than 1 value means the model improves the performance of the baseline, smaller than 1 value means the baseline outperforms the model. For the averaged metric, all the top performing text embeddings outperforms the baseline with a considerable gap (SimCSE is ranked 101th place). However, the improvements across different individual tasks are heavily imbalanced: 
\begin{itemize}
    \item Classification tasks: The logistic regression classifier, with a maximum of 100 iterations, is trained using the train set embeddings and its performance is evaluated on the test set \cite{muennighoff2022mteb}. It can be seen from Table \ref{tab2} that all of these top 25 best performing models are better than the baseline method SimCSE with varied improvements between 9\% and 21\%. 
    \item Clustering tasks: A mini-batch k-means model is trained on the embedded texts, utilizing a batch size of 32 and setting k to match the total number of unique labels with the v-measure as the metric \cite{muennighoff2022mteb}. 
    All of the top performing text embedding models outperform the baseline by around 35\%-57\% increase over the baseline performance. 
    \item Pair Classification (Pair-C): Duplicate or paraphrase pairs with binary labels are embedded and  the average precision score based on cosine similarity on text embeddings is used as the main metric \cite{muennighoff2022mteb}.  
    The performance of all the top performing text embedding models is superior (with varied improvements between 15\% and 20\%) to the baseline in the Pair Classification task.
    
    \item Reranking tasks: Given a query and a list of relevant and irrelevant reference texts, cosine similarity is used to compare the embeddings and rank the references with MAP being the main metric \cite{muennighoff2022mteb}. 
    The Reranking tasks show an improved performance (between 22\% and 28\%) from all MTEB leading text embedding models compared to the baseline.
    
    \item Retrieval tasks: Given a corpus, queries and a mapping for each query to relevant documents from the corpus, cosine similarity scores on the embeddings between query and documents are used to rank documents for each query, with  nDCG@10 being the main metric \cite{muennighoff2022mteb}. 
    The most considerable enhancement in the top-rated text embedding models of MTEB  is observed in Retrieval tasks, with the majority of these models more than doubling the performance of baseline model. 
    
    \item Semantic Textual Similarity (STS) tasks: Given sentence pairs labeled with continuous scores with higher numbers indicating more similar sentences, Spearman correlation based on cosine similarity between sentence pair embeddings is main metric \cite{muennighoff2022mteb}. 
   The increase in performance is moderate in STS tasks compared to other tasks for all top performing MTEB text embedding models, with the best performing model increasing 8.4\% over the baseline performance.
   
    \item Summarization tasks: Given human-written and machine-generated summaries, cosine similarity between embeddings of machine summary and human summary is used  to score the machine summaries with Spearman correlation being the main metric \cite{muennighoff2022mteb}. Unlike other tasks, most of the top performing text embedding models are not able to outperform the baseline performance on summarization tasks. 
\end{itemize}

From the results in Table \ref{tab2}, it can be seen that compared to the baseline text embedding SimCSE published in 2021, most the top 25 best performing MTEB text embedding models (mostly published in 2023 and 2024) are not remarkably better than the baseline on all tasks, especially on Summarization tasks. All the top 25 text embedding models are notably
better than the baseline model on Retrieval, Reranking, Clustering and Pair Classification tasks, especially on Retrieval task. The proposed methodologies appear to primarily impact the performance of retrieval tasks. However, it might be related to the training and fine-tuning datasets used by the top performing models and their similarity to MTEB benchmarks. Popular datasets used by the top performing models include StackExchange, Reddit, S2ORC,  NLI \cite{nligao2021simcse}, FEVER \cite{thorne2018fever}, NQ \cite{nq1karpukhin2020dense,nq2kwiatkowski2019natural}, HotpotQA \cite{yang2018hotpotqa}, Quora \cite{iyer2017quora}, MSMARCO, etc.  These datasets are similar to MTEB benchmark datasets especially on Retrieval and Clustering tasks. Apart from datasets similarity, there are many efforts made by the state of the art embeddings to deal with the asymmetric tasks such as Retrieval, including generation of more synthetic asymmetric datasets as in \cite{e5mistralwang2023improving,e5instructwang2024multilingual}, instruction based embeddings as in \cite{echoispringer2024repetition,e5instructwang2024multilingual,e5mistralwang2023improving,llm2vec,gritmuennighoff2024generative,lee2024gecko,bgexiao2023c}, asymmetric formatting as in \cite{lee2024gecko} and so on.  Generally speaking, The results from Table \ref{tab2} show that: the overall performance on MTEB benchmark are improved considerably by recent advances in universal text embeddings especially on Retrieval tasks while the performance on Summarization task sees no notable improvement compared to the baseline method.  

In terms of universality on languages, most of these models are trained on specific languages, typically English, and do not inherently accommodate multilingual data. This lack of language universality restricts their application in global, multilingual contexts. In the work of \cite{e5instructwang2024multilingual,e5mistralwang2023improving}, the authors use proprietary LLMs to generate synthetic data for a diverse range of text embedding tasks in 93 languages, covering hundreds of thousands of embedding tasks, which shows good performance on high-resource languages. However, for low-resource languages, there is still room for improvement as current open-source LLMs are not adequately pre-trained on them.  In terms of the universality on text length, MTEB has Sentence to Sentence (S2S) tasks as well as Paragraph to Paragraph (P2P) tasks where the former only compare titles, while the latter include both title and content \cite{muennighoff2022mteb}. For clustering tasks, Arxiv, Biorxiv, Medrxiv, Reddit and StackExchange have both S2S and P2P version, where S2S tasks have short texts with on average 57-115 chars and P2P tasks have long texts with on average 728-1981 chars. Most top performing text embeddings have better performances on P2P tasks on  Arxiv, Biorxiv, Medrxiv, Reddit. However, on StackExchange data, most top performing text embeddings have much better  performance on S2S tasks. This might be more related to the informativeness nature of datasets instead of to the text length. Better benchmark datasets design related to text length is needed. For example, comparing the clustering performance on long text data before and after different extends of summarization could be an option.

\subsection{Model efficiency analysis}

\begin{figure*}[t!]
\includegraphics[width=1\textwidth]{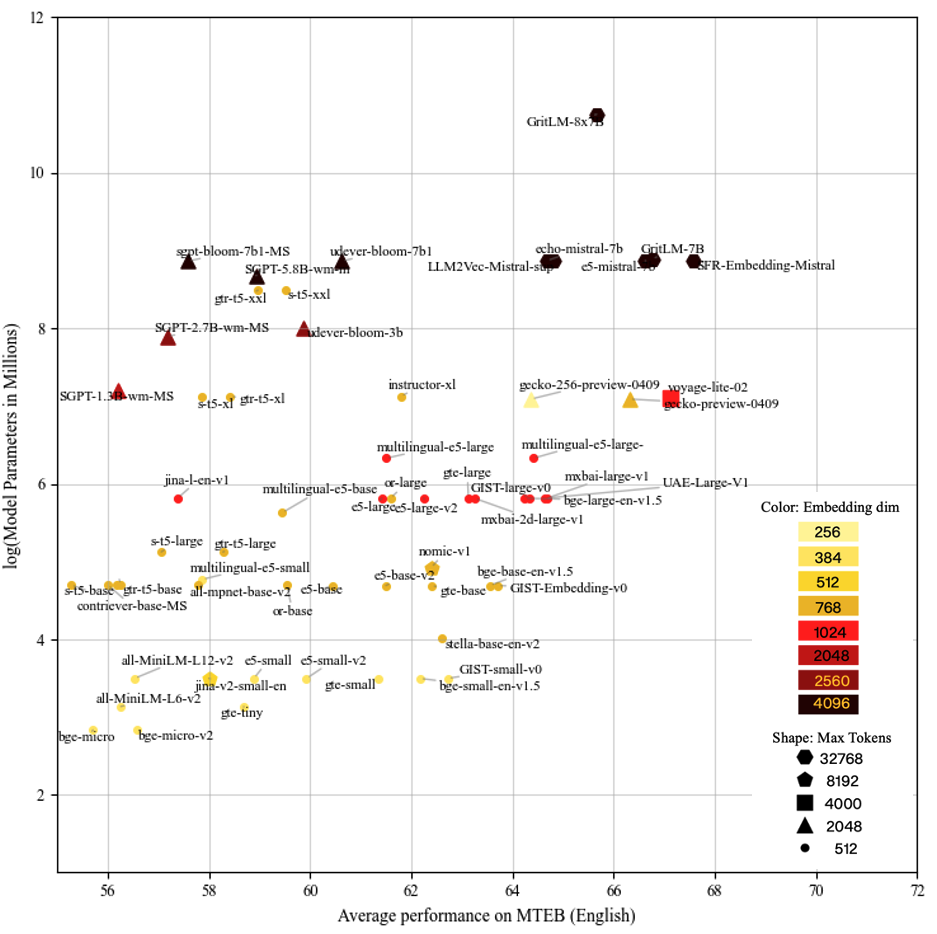}
\caption{The top performing text embeddings on MTEB benchmark: X-axis is the average performance over 56 MTEB benchmark datasets, Y-axis is the log of Model parameter numbers (in Millions). Different colors indicate different embedding dimensions and different shapes indicate different max token sizes. } \label{fir}
\end{figure*}

In the field of AI and NLP, Occam's Razor could be applied in the process of comparing algorithms or models. If two models perform similarly well, the principle would suggest opting for the simpler one, as it is likely to be more efficient and less prone to overfitting. To compare the efficiency of different text embedding models, the average performance on MTEB English benchmark of  state of the art text embedding models and their corresponding model parameters (log wise) are plotted in Figure \ref{fir}. The efficiency of the downstream tasks using text embedding as input is related to the dimension of the embeddings. Larger embedding dimension indicates higher computational cost, storage/memory cost and latency for downstream tasks. Hence, the embedding dimension for each model is also illustrated in Figure \ref{fir}, with varying colors denoting different dimensions. The spectrum ranges from light yellow (representing a dimension of 256) to deep red (representing a dimension of 4096). The max token size which is related to the model efficiency when dealing with long input texts is illustrated by different shapes in Figure \ref{fir} with: small circle (512/514 max input tokens), triangle (2048 max input tokens), square (4000 max input tokens), pentagon (8192 max input tokens), hexagon (32768 max input tokens). 

\textbf{Model sizes:}  In previous studies \cite{muennighoff2022mteb}, it was found that the performance strongly correlates with model size, which can be  identified in Figure \ref{fir}. For example, when the parameters of SGPT increases from 1.3B to 5.8B, the performance increases from 56.2\% to 58.93\%. Such kind of scaling behavior encourages many studies to scale model size up in order to provide state of the art results across different  embedding tasks. Recently, there are more and more models focus on generating text embeddings from LLMs  because it does not need the contrastive pre-training step used in existing state of the art text embedding models as LLMs are extensively pre-trained on web-scale data already \cite{e5instructwang2024multilingual,gritmuennighoff2024generative,llm2vec,SFRAIResearch2024,echoispringer2024repetition,lee2024gecko}. 
However, LLMs are computationally expensive, resource-intensive, and difficult to deploy in real-world applications, particularly on devices with limited processing power. Moreover, the marginal gains in performance do not always justify the substantial increase in parameter size, complexity and resource requirements.  Additionally, we can see that when GritLM 7B is scaled up to GritLM 7x8B, the overall performance on MTEB benchmark decreases across all tasks (note that Grit models are both embedding and generation models). The performances of 7B parameters models  vary a lot from 57.59\% (sgpt-bloom-7b1-msmarco) to 67.56\% (SFR-Embedding-Mistral) as shown in Figure \ref{fir}, while jina-embeddings-v2-small-en achieves better performance than sgpt-bloom-7b1-msmarco with only 33M parameters. Furthermore, the two 1.2B models voyage-lite-02-instruct and google-gecko.text-embedding-preview-0409 demonstrate comparable or superior performances to most 7B LLMs based models, which suggests that there is significant room for enhancement in the efficiency of numerous state of the art text embedding models.

\textbf{Embedding sizes:} Deploying text embedding involves two steps: a constant forward pass to compute the embedding, and its use for downstream applications \cite{sato2021vertex,varma2019extreme}. The computation costs for the second step rise with the embedding dimensionality, data size, and label space, which can exceed the feature computation cost for large scale systems \cite{dean2009challenges,sun2017revisiting}. In some RAG systems where documents are stored as text embedding vectors, the embedding dimension is also related to the storage and memory cost, especially for large scale RAG systems.  The top-performing text embedding dimension sizes vary from 256 to 4096, while the largest embedding dimension reported in MTEB English benchmark is 12288 from text-similarity-davinci-001 and text-search-davinci-001. MRL \cite{mrlkusupati2022matryoshka} and 2dMSE \cite{2dmrlli20242d} propose new loss functions to allow  first $m$ dimensions of the embedding to be independently capable of being a general purpose text embedding too. Among the top performing text embedding models,  Gecko \cite{lee2024gecko} embeddings are the most compact with google-gecko.text-embedding-preview-0409 (768 dimensions) and google-gecko-256.text-embedding-preview-0409 (256 dimensions).

\textbf{Max token sizes:} The max token size limits the length of the input text to be embedded. When the input length exceeds the max token size,  the most straightforward solution is to truncate the input text to the maximum allowed length.
However, this approach has the drawback of eliminating potentially relevant text. An alternative strategy involves partitioning the input text into smaller chunks, embedding each chunk separately,  then combining the embeddings of all chunks. Although this method preserves the entirety of the input text, it reduces the efficiency of the embedding model.
The max token sizes of top performing text embedding models in MTEB English benchmark vary from 512 to 32768. For BERT like based models, their  max token size is usually 512, while text embedding models based on Mistral-7B have the max token size of 32768. To be noted that different LLMs may have different max token sizes. For example, Llama 2 \cite{touvron2023llama2} with 7B, 13B, and 70B parameters have a max token size of 4096. Further more, the max token size can be extended in various ways for both LLMs \cite{extendzhang2024soaring} and BERT like models \cite{nussbaum2024nomic}. As MTEB lacks datasets with larger length, it is not clear how Max token sizes may impact the performance of universal text embedding models.


\subsection{Limitations}
Apart from the limitations analyzed above in the previous sections, several other limitations are identified in this section: 

\paragraph{\textbf{Data:}} The complexity of comparing different models arises due to variations in numerous factors such as training data, back-bone model, loss function, training strategy, negative-sampling strategy, embedding dimension, among others. It is challenging to establish a fair comparison due to these differences. Few papers analyze the similarity between their training, pre-training or fine-tuning data and the MTEB benchmark datasets which makes it unclear whether MTEB test datasets are in-domain, partially in-domain or out-of-domain for these text embedding models. Many studies claim that the dataset diversity is important to achieve the universal text embeddings \cite{bgexiao2023c,gteli2023towards,e5instructwang2024multilingual,lee2024gecko}. However, the current literature lacks a metric to accurately measure this dataset diversity, further complicating the issue. This gap in the literature underscores the need for a more rigorous approach to assessing dataset diversity in future studies.

\paragraph{\textbf{Instruction:}} Instruction refers to the task instruction, which specifies a description of the task that the embedding will be used for (as shown in Equation \ref{inst}) in order to build universal text embedding models that can generalize across a large variety of tasks \cite{echoispringer2024repetition,e5mistralwang2023improving,asai2022task}. Many studies have shown that adding instructions has a considerable impact on the performance. However, there are several limitations. Firstly, the effectiveness of the instruction is highly dependent on its quality and specificity. If the instruction is vague or ambiguous, the model may fail to embed the text properly, leading to poor performance on the task. Additionally, creating precise and comprehensive instructions for every possible task can be a labor-intensive and time-consuming process. Secondly, the model's ability to interpret and follow the instructions is limited by its current understanding of language, which may not perfectly align with human understanding. This could lead to misinterpretations and errors. Furthermore, the incorporating instructions into text embeddings increases the input length which can be computationally intensive, particularly for large datasets and large models. Finally, 
few papers explain how instruction impacts the text embedding for symmetric and asymmetric tasks and helps improve the performance theoretically. How out-of-domain instructions impact the model performance is not clear neither.

\paragraph{\textbf{Benchmark:}} Massive Text Embedding Benchmark (MTEB) is the most popular and used benchmark for universal text embeddings.  There are several already identified limitations of MTEB including: lacking long texts datasets (most test datasets MTEB have fewer than 500 chars), task imbalance (15 datasets on Retrieval task, 12 datasets on Classification task while only 1 dataset for Summarization task), limited multi-languages evaluation datasets and no programming language (code) datasets \cite{muennighoff2022mteb}.   
Understanding syntax thoroughly is essential for a text embedding model to accurately determine the relationships between words, which aids in achieving a level of language comprehension that mirrors human cognitive processes \cite{zhang2023well}. The capacity of text embedding models to generalize across various syntactic contexts is not sufficiently examined in the existing benchmark. Therefore, to evaluate the proficiency of text embedding models in understanding syntax, it would be beneficial to incorporate more datasets that focus on syntactic aspects. 
The variety of datasets can certainly be enhanced. For instance, out of the 11 datasets used for clustering in MTEB, six originate from scientific articles published on platforms like Arxiv, Biorxiv, and Medrxiv. It would be beneficial to include datasets from different fields like finance, business, arts, culture, health, travel, and more to broaden the scope.

\paragraph{\textbf{Similarity measures}}
Distance metrics $d(\cdot, \cdot)$ in vector spaces must obey certain axioms or geometric constraints \cite{cao2019random,cao2019random1} including:
\begin{itemize}
    \item Reflexivity: $d(\mathbf{x}_i, \mathbf{x}_i) = 0 $ 
    \item Nonnegativity: $d(\mathbf{x}_i, \mathbf{x}_j) \geq 0 $ 
    \item Symmetry: $d(\mathbf{x}_i, \mathbf{x}_j) = d(\mathbf{x}_j, \mathbf{x}_i) $
    \item Triangle inequality: $d(\mathbf{x}_i, \mathbf{x}_k) \leq d(\mathbf{x}_i, \mathbf{x}_j) + d(\mathbf{x}_j, \mathbf{x}_k) $
\end{itemize}
Cosine similarity is widely used in the literature and MTEB benchmark to measure similarity between text embeddings, which also obeys symmetry and an analogue of the triangle inequality \cite{griffiths2007topics}. 
However, psychological representations of similarity do not always obey these constraints. The authors from \cite{tversky1977features,tversky1986nearest} show that some important aspects of human judgments of item similarity can not be captured by some of the geometric axioms of vector spaces.  
Researchers from  \cite{peterson2020parallelograms,tversky1977features} demonstrate that  human relational similarity judgments violate the geometric constraints of symmetry and the triangle inequality. 
A famous example in terms of violation of symmetry is that people judge North Korea to be more similar to China than the other way around \cite{peterson2020parallelograms}. 
Furthermore, the authors from \cite{steck2024cosine} conclude that cosine-similarity can yield arbitrary and meaningless similarities. Compared to the term of distance or kernel, dissimilarity and similarity are more general terms, which do not have the constraints to be a metric or positive semi-definite \cite{pekalska2005dissimilarity,cao2021novel}. New (dis)similarity measures that aligns better with human judgments could be an interesting and important future directions.

\section{Conclusions}
In this article, an overview of the recent advances in universal text embedding models is provided. Various definitions of universal text embeddings from the literature are integrated in this work: universal text embedding is a unified comprehensive text embedding model that can address a multitude of input text length, downstream tasks, domains and languages. The top performing universal text embedding models on MTEB benchmark are categorized into three groups:  data focus, loss function focus and LLM focus. Representative works of each category are presented and compared. These state of the art methods have made significant improvements and innovations in terms of training data quantity, quality and diversity; synthetic data generation for universal text embeddings as well as using large language models as backbones.  The overall performance on MTEB English benchmark are remarkably improved by these recent universal text embedding models especially on Retrieval, Reranking, Clustering and Pair Classification tasks. 

However,  there remains a significant gap that needs to be addressed in the current state of the art universal text embedding models. First of all, unlike the considerable improvements on Retrieval tasks, little improvement is made by current state of the art solutions on summarization tasks. Secondly, most of existing text embeddings are trained on specific languages, typically English, and do not inherently accommodate multilingual data. This lack of language universality restricts their application in multilingual contexts. Thirdly, current benchmarks lack domain diversity. Datasets from different fields like finance, business, arts, culture, health, travel, and more with diverse text lengths should be included to broaden the scope and test the domain generalization ability of universal text embedding models.

In terms of future research, there are numerous broad areas that merit further exploration. One such area is the construction of a more comprehensive and diverse benchmark that can test the universality  holistically across domains, tasks, input lengths and languages. The redundancy of the benchmark datasets should be minimized to reduce the computational cost of testing. Secondly, developing solutions to make universal text embeddings more sustainable and cost-effective in terms of training, inference and downstream tasks usage is also an interesting direction.  Additional future research could focus on in-depth understanding on instructions, its impact on symmetric and asymmetric tasks, its generalization ability and so on. Finally, another interesting future direction could be proposing novel (dis)similarity measures that can produce human-like asymmetries from vector-space text embeddings.




\vskip 0.2in
\bibliographystyle{ieeetr}
\bibliography{sample}
\end{document}